\documentclass {article}

\usepackage{graphicx}

\begin{document}

\title{Charged Holostars}

\author{Michael Petri\thanks{email: mpetri@bfs.de} \\Bundesamt f\"{u}r Strahlenschutz (BfS), Salzgitter, Germany}

\date{{June 16, 2003 (v1)} \newline {May 1, 2004 (v3)}}

\maketitle

\begin{abstract}
In a recent paper the so called holographic solution, in short
holostar, was introduced as a new spherically symmetric solution
of Einstein field equations. This paper extends the holostar
solution to the charged case.

A charged holostar is an exact solution of the Einstein field
equations with zero cosmological constant and an interior string
type matter. It has properties similar to the known
Reissner-Nordstr\"{o}m (RN) black hole solution. The exterior
metrics and fields of both solutions are equal . In contrast to a
RN black hole, a charged holostar has a singularity free interior
matter distribution $\rho = 1 / (8 \pi r^2)$ with an overall
string equation of state: $P_r = -\rho$ and $P_\perp =0$. Similar
to the uncharged holostar solution, the charged holostar has a
spherical boundary membrane consisting out of tangential pressure,
but no mass-energy. The boundary is situated roughly a Planck
coordinate distance outside of the outer horizon of the
RN-solution.

The geometric properties of the charged holostar solution are very
conveniently described in terms of the so called geometric mass
$M_g = M + r_0/2$. $r_0$ is a Planck size correction to the
gravitational mass $M$ with $r_0 \approx 2 r_{Pl}$. The geometric
mass of a charged holostar is always larger than its charge in
natural units. For a large holostar this condition is practically
identical to the classical condition $M \geq Q$. Whereas RN
solutions with $M < Q$ are possible in principle, and are excluded
from the physically acceptable solution space by the cosmic
censorship hypothesis, a charged holostar with $M_g > Q$ doesn't
exist.

The total exterior charge $Q$ of a holostar can be attributed to
the charge of its massive interior particles. $Q$ is derived by
the proper integral over the interior charge density. The interior
mass-energy density $\rho$ splits into an electromagnetic
contribution $\rho_{em}$ and a "matter" contribution. Both
contributions are proportional to $1/r^2$. Yet the {\em total}
interior mass-energy density and the total principal pressures are
exactly equal to the uncharged case. The same is true for the
interior metric.

The ratio of electro-magnetic to total energy density
$\rho_{em} / \rho = 4 \pi Q^2/A$ is {\em constant} throughout the
whole interior. It is related to the dimensionless ratio of the
exterior conserved quantities $Q^2$ and $A$ (or alternatively
$Q/M_g$). An extremely charged holostar has a surface area $A = 4
\pi Q^2$, so that its interior energy density consists entirely
out of electromagnetic energy. The tangential pressure in the
membrane of an extremely charged holostar vanishes, so that the
energy-density is continuous at the boundary.

Similar to the uncharged holostar, the charged holostar solution
admits charged "particle-like" solutions with nearly zero
gravitational mass, but with appreciable charge in natural units.
A charged particle with mass and charge comparable to the
electron, however with zero spin, is a genuine solution.

The equations for a zero mass extremely charged holostar are
extrapolated to the rotating case. Under the assumption, that the
electron can be identified with an extremely charged holostar of
minimal mass, a scaling law for the fundamental area $r_0^2$ is
conjectured, according to which $r_0^2 \approx 4 \hbar$ at the
Planck-energy and $r_0^2 \simeq \sqrt{3/4} \, \, 4 \hbar$ at the
low energy scale.

A large holostar can be regarded as the classical analogue of a
large loop quantum gravity (LQG) spin-network state, if one
identifies the links of the LQG spin-network state with the
interior massless particles of the classical holostar solution.
The Barbero-Immirzi parameter is determined to be equal to $\gamma
= \sigma /(\pi \sqrt{3})$, where $\sigma$ is the mean entropy per
ultra-relativistic particle. $\gamma$ is larger by a factor of
$\approx 4.8$ than the LQG-result. An explanation for the
discrepancy is given.

An (approximate) entropy conservation law for self gravitating
systems in general relativity is proposed.

\end{abstract}

\section{\label{sec:intro}Introduction}

In \cite{petri/bh} a new class of solutions to the spherically
symmetric field equations of general relativity was derived. The
solutions are characterized by an interior non-zero string type
matter-density and a boundary membrane with zero energy density,
but non-zero tangential pressure. The interior string type
pressure generally is anisotropic.

One of the new solutions, the so called "holographic" solution, in
short holostar, was discussed in greater detail in
\cite{petri/hol, petri/thermo}. It's interior equation of state is
that of a spherically symmetric string vacuum, bounded by a
two-dimensional membrane. The membrane's pressure is exactly equal
to the pressure attributed to the (fictitious) membrane of a black
hole according to the membrane paradigm \cite{Thorne/mem}.
Therefore the exterior properties of the holostar are guaranteed
to be virtually indistinguishable from the properties of a black
hole. The holostar's entropy and temperature are equal to the
Hawking result up to a constant factor \cite{petri/thermo}. Yet
the holostar has no event-horizon and to singularities. It appears
to be an amazingly self-consistent model for the most compact,
self-consistent static solution of the Einstein field equations
that is not a black hole.

So far only uncharged solutions were discussed. In this paper I
attempt to generalize the previous results to the case of a
charged self gravitating body. Although the practical
applicability of the charged holostar solution is expected to be
limited, as self gravitating objects of astrophysical interest are
assumed to be essentially uncharged, the charged holostar solution
is of considerable interest from a theoretical point of view: From
the study of black holes it is well known that the properties of
the charged black hole solutions are in many respects similar to
the properties of the spinning black hole solutions. In
gravitational collapse of large stars one expects that a highly,
almost maximally spinning black hole is formed. Observational
evidence for a black hole with high angular momentum ($a/M \approx
0.95$) is given in \cite{Fabian}. Therefore a spinning holostar
solution is of considerable interest and high astrophysical
relevance. It might be possible to infer some of the properties of
a spinning holostar from the charged solution.

\section{\label{sec:field_equations}Field equations for a spherically symmetric charged system}

The approach taken in this paper is similar to the route taken in
\cite{petri/bh}. As a basis for the derivation the well
known Schwarzschild coordinate system in the (+ - - -)
sign-convention with units $ c = G = 1$ is used. Without loss of
generality the metric of a static spherically symmetric
space-time, charged or uncharged, can be expressed  in the
following form:

\begin{equation} ds^2 = B(r) dt^2 - A(r) dr^2 - r^2 d\theta^2 - r^2 \sin^2 \theta d\varphi^2
\end{equation}

The charge distribution $\rho_{c}$, the electromagnetic energy
density $\rho_{em}$ and the "matter fields", i.e. the
(non-electromagnetic) mass-density $\rho_m$ and its principal
pressures ${P_r}_m$ and ${P_\theta}_m$, are spherically symmetric.
They only depend on the radial coordinate value $r$. Whenever
appropriate the functional dependence of the relevant quantities
on $r$ will not be written out explicitly.

The stress-energy tensor $T_\mu^\nu$ for a charged holostar is
given by the sum of a "matter-" and an
electromagnetic-component.\footnote{Note, that according to common
knowledge charge is always associated with mass-energy. Therefore
the distinction between a "matter" and an electromagnetic
component in the stress energy tensor should not be viewed as a
statement that the two components are necessarily physically
separable, but rather as a mathematical idealization which is
useful for the calculations.} When the term "matter" component is
used throughout this paper, I mean the components of the
stress-energy tensor of the self gravitating object which are not
electromagnetic in origin. With this distinction the matter
component of the stress-energy tensor can be expressed as:

\begin{equation} \left(T_\mu^\nu \right)_{m} = \left( \begin{array} {cccc}
\rho_m(r) \\
& -{P_r}_m(r) \\
&& -{P_\theta}_m(r) \\
&&& -{P_\theta}_m(r) \\
\end{array} \right)
\end{equation}

Its trace $T_m$, which is in general non-zero, is given by:

\begin{equation} T_m = (T_\mu^\mu)_m = \rho_m -{P_r}_m - 2 {P_\theta}_m
\end{equation}

For a spherically symmetric system the electromagnetic component
of the stress-energy tensor is given by the following, traceless
tensor:

\begin{equation} \label{eq:Tem}
\left(T_\mu^\nu \right)_{em} = \left( \begin{array} {cccc}
\frac{E^2(r)}{8 \pi} \\
& \frac{E^2(r)}{8 \pi} \\
&& -\frac{E^2(r)}{8 \pi} \\
&&& -\frac{E^2(r)}{8 \pi} \\
\end{array} \right)
\end{equation}

$E(r)$ is the electromagnetic\footnote{Because of spherical
symmetry we are dealing with an electro-static problem. Therefore
"electromagnetic" in the context of this paper always means
"electrostatic"} field strength at radial position $r$, in natural
units.

Note that the interior string-type stress-energy
tensor of the uncharged holostar solution with

$$T_\mu^\nu = diag(\rho , \rho, 0, 0)$$

i.e. with $P_r = -\rho$ and $P_\theta = P_\varphi = 0$, can be
constructed from the sum of a vacuum contribution

$$\left(T_\mu^\nu\right)_{vac} = \rho_{vac} \, diag(1 , 1, 1, 1)$$

and an electromagnetic contribution

$$\left(T_\mu^\nu\right)_{em} = \rho_{em} \, diag( 1 , 1 , -1, -1 )$$

if the vacuum energy-density $\rho_{vac}$ and the electromagnetic
energy density $\rho_{em} = E^2 / (8 \pi)$ are equal. We will see
later, that the total interior stress energy-tensor of the charged
holostar solution is identical to the uncharged case. The charged
and the uncharged solution only differ in their exterior fields.

The electromagnetic energy density must not necessarily
be associated with a net charge. The electromagnetic energy
density depends on the fields squared, therefore it is possible to
have $\overline{E^2} \neq 0$, even if $\overline{Q} = 0$.

In the case of a static spherically symmetric charge distribution
one can express the magnitude of the radially symmetric electric
field $E(r)$ in terms of the total charge $Q(r)$ enclosed in a
concentric region bounded by the radial coordinate $r$:

\begin{equation} \label{eq:E(r)}
E(r) = \frac{Q(r)}{r^2}
\end{equation}

For the Reissner-Nordstr\"{o}m solution the electromagnetic field
tensor (\ref{eq:Tem}), with $E(r)$ given by equation
(\ref{eq:E(r)}), is known to be valid outside the horizon, i.e. in
the vacuum region where $Q = Q(r_h)$ is constant. In this case it
is easy to evaluate $Q$ by a Gaussian flux integral. Let us make
the assumption, that equations (\ref{eq:Tem}, \ref{eq:E(r)}) also
hold {\em inside} a spherically symmetric self gravitating body, and
that $Q(r)$ is given by the proper integral over the interior
charge distribution:\footnote{Because charge is a relativistic
invariant, the integral must be taken over the proper volume
element. It is not possible to integrate over the improper (flat)
spherical volume element, such as in the determination of the
gravitational mass.}

\begin{equation} \label{eq:Q(r)}
Q(r) = \int_0^r{\rho_{c}(r) dV} = \int_0^r{\rho_{c}(r) 4 \pi r^2
\sqrt{A} dr}
\end{equation}

$\rho_c$ is the interior charge density. It should not be confused
with the energy density $\rho_{em} = E^2 / (8 \pi)$ of the
electromagnetic field.

Whenever the stress-energy tensor of the space-time is known, it
is convenient to write the field equations in the following form:

\begin{equation} \label{eq:R=8piT}
R_{\mu \nu} = -8 \pi \left( {T_{\mu \nu} - \frac {T} {2}}
\right)
\end{equation}

with

$$T_{\mu \nu} = \left(T_{\mu \nu} \right)_{m} + \left(T_{\mu \nu} \right)_{em} $$

The components of the Ricci-tensor on the left side of equation
(\ref{eq:R=8piT}) can be calculated from the metric coefficients
and their first and second derivatives. The actual expressions can
be found in any textbook.\footnote{see for example \cite[p.
300]{Weinberg/GR} or \cite[p. 128, 226]{Flie/AR}} The right side
of equation (\ref{eq:R=8piT}) is easily evaluated from the
expressions for $(T)_{m}$ and $(T)_{em}$ given above. Only the
diagonal components of the field equations give non-zero
expressions. The equation for $R_{\varphi\varphi}$ is a trivial
multiple of the equation for $R_{\theta\theta}$. Thus we are left
with three equations:

\begin{equation} \label{eq:R00} R_{tt}=-\frac{B''}{2A} + \frac{B'}{4A} \left(
{\frac{A'}{A} + \frac{B'}{B}} \right) - \frac{B'}{rA} = - 4 \pi B
(\rho_m + {P_r}_m + 2 {P_\theta}_m + \frac{Q^2}{4 \pi r^2} )
\end{equation}

\begin{equation} \label{eq:R11} R_{rr}=\frac{B''}{2B} - \frac{B'}{4B} \left(
{\frac{A'}{A} + \frac{B'}{B}} \right) - \frac{A'}{rA} = - 4 \pi A
(\rho_m + {P_r}_m - 2 {P_\theta}_m - \frac{Q^2}{4 \pi r^2} )
\end{equation}

\begin{equation} \label{eq:R22} R_{\theta\theta}=-1 - \frac{r}{2A} \left(
{\frac{A'}{A} - \frac{B'}{B}} \right) + \frac{1}{A} = - 4 \pi r^2
(\rho_m - {P_r}_m + \frac{Q^2}{4 \pi r^2} )
\end{equation}

The field equations for the charged case can be deduced from the
field equations of the uncharged case, simply by making the
following replacements:

\begin{equation}
\rho \rightarrow \rho_{tot} = \rho_m + \frac{E^2}{8 \pi}= \rho_m +
\frac{Q^2}{8 \pi r^4}
\end{equation}

\begin{equation}
P_r \rightarrow {P_r}_{tot}  =  {P_r}_m - \frac{E^2}{8 \pi}=
{P_r}_m - \frac{Q^2}{8 \pi r^4}
\end{equation}

\begin{equation}
P_\theta \rightarrow {P_\theta}_{tot} =  {P_\theta}_m +
\frac{E^2}{8 \pi}= {P_\theta}_m + \frac{Q^2}{8 \pi r^4}
\end{equation}

Keep in mind, that with the notation in equations (\ref{eq:R00},
\ref{eq:R11}, \ref{eq:R22}), $\rho_m$, ${P_r}_m$ and
${P_\theta}_m$ exclusively refer to the matter-contribution,
excluding the electro-magnetic contribution. For the gravitational
field, or rather for the metric, only the total
mass/energy-density and the total principal pressures are
relevant. Whenever the total values are referenced, they will
appear without subscript throughout this paper.

\section{General properties of the solution}

By multiplying equations (\ref{eq:R00}) and (\ref{eq:R11}) with
the metric coefficients $A$ and $B$ respectively and summing up,
the following expression results:

\begin{equation} \label{eq:R00A+R11B}
R_{00} A + R_{11} B = -\frac{B}{r}\left(\frac{A'}{A} +
\frac{B'}{B}\right) = - 8 \pi A B (\rho_m + {P_r}_m)
\end{equation}

which can be reformulated as:

\begin{equation} \label{eq:lnAB}
\left(\ln{A B}\right)' = 8 \pi r A (\rho_m + {P_r}_m)
\end{equation}

Equation (\ref{eq:lnAB}) is identical to the uncharged case.

For a spherically symmetric electric field we have

\begin{equation} \label{eq:rho+Pr:em}
\rho_{em} + {P_r}_{em} = 0
\end{equation}

Therefore the electromagnetic field does not contribute to $\rho +
P_r$, so that the sum of the total energy density and total radial
pressure is equal to the sum of the matter-components only:

\begin{equation} \label{eq:rho+Pr:tot}
\rho + P_r = \rho_{m} + {P_r}_{m}
\end{equation}

With equation (\ref{eq:R00A+R11B}) the term $B'/B$ can be
eliminated in equation (\ref{eq:R22}), giving the following
differential equation for the radial metric coefficient $A$:

\begin{equation} \label{eq:r/A}
\left( \frac{r}{A} \right)' = 1 - 8 \pi r^2 \left( \rho_m +
\frac{Q^2}{8 \pi r^4} \right)
\end{equation}

Equation (\ref{eq:r/A}) differs from the respective formula for
the uncharged case by adding the term $Q^2 / (8 \pi r^4)$ to the
mass-energy density of the matter contribution, $\rho_m$. The
added term is nothing else than the electromagnetic energy density
$\rho_{em} = E^2 / (8 \pi)$.

The tangential pressure ${P_\theta}_m$ can be calculated via one
of the equations (\ref{eq:R00}, \ref{eq:R11}):

\begin{equation} \label{eq:Pt}
{P_\theta}_m + \frac{E^2}{8 \pi} = \frac{1}{8 \pi A B}\left(
\frac{B''}{2} + \frac{B'}{r}\right) - \left( \frac{\rho_m +
{P_r}_m}{4}\right)\left(\frac{rB'}{B}+2 \right)
\end{equation}

Alternatively the tangential pressure can be derived from the
continuity equation. In the general case $A B \neq 1$ this is
usually computationally much less involved, especially if the
total stress energy tensor, i.e.  the sum of electromagnetic and
matter contributions, is known and has a simple form.

\begin{equation} \label{eq:Pt:cont}
{P_\theta} = {P_\theta}_m + \frac{E^2}{8 \pi} = P_r +
\frac{r {P_r}' }{2}+ \frac{r B'}{B}\left(\rho +
P_r \right)
\end{equation}

Whenever $\rho_m$, ${P_r}_m$ and the electromagnetic field $E$ are
known, the metric coefficients $A$ and $B$ and ${P_\theta}_m$ can
be determined by the following procedure:

\begin{itemize}
\item Integrate equation (\ref{eq:r/A}) to obtain the radial
metric coefficient $A$; in order to do this $\rho_m$ and $E^2/(8
\pi)$ must be known

\item Determine $B$ by integrating equation (\ref{eq:lnAB}); the
integration requires knowledge of $\rho_m$ and ${P_r}_m$ ($A$ has
been obtained in the first step)

\item Determine the tangential pressure by equation (\ref{eq:Pt})
or (\ref{eq:Pt:cont}); this requires knowledge of $E$, $\rho_m$,
${P_r}_m$ (and of $A$ and $B$, which were obtained in the previous
two steps)
\end{itemize}

On the other hand, if the metric is known, the matter-fields
$\rho_m$, ${P_r}_m$ and ${P_\theta}_m$ can be determined by
differentiation of the metric coefficients. This, however,
requires a prior knowledge of the electromagnetic energy-density
$E^2 / (8 \pi)$. We find:

\begin{equation} \label{eq:rho}
\rho = \rho_m + \frac{E^2}{8 \pi} = \frac{1}{8 \pi r^2}\left(
1-\left( \frac{r}{A}\right)' \right)
\end{equation}

\begin{equation} \label{eq:Pr}
P_r + \rho = {P_r}_m + \rho_m = \frac{\left( \ln{A
B}\right)'}{8 \pi r A}
\end{equation}

\begin{equation} \label{eq:Pt2}
P_\theta  = {P_\theta}_m+ \frac{E^2}{8 \pi} = (\ref{eq:Pt}) \,
{or} \, (\ref{eq:Pt:cont})
\end{equation}

Equations ( \ref{eq:rho}, \ref{eq:Pr}, \ref{eq:Pt2}) carry a very
important message: $\rho$, $P_r$ and $P_\theta$, which appear on
the left side of the above equations, are the diagonal components
of the {\em total} stress-energy tensor for a static charged
spherically symmetric system, {\em including} the electromagnetic
contribution. The values of $\rho$, $P_r$ and $P_\theta$ can be
determined exclusively from the metric: First $\rho$ is determined
via equation (\ref{eq:rho}). Knowing $\rho$ one obtains $P_r$ from
equation (\ref{eq:Pr}) and last $P_\theta$ via equation
(\ref{eq:Pt2}). In order to determine the total stress-energy
tensor of a spherically symmetric, charged self gravitating object
we therefore don't need to know anything about its electromagnetic
field. How the electromagnetic field contributes to the total
mass-energy of the space-time, be it just a fraction, be it zero
or large, is irrelevant. In order to determine the {\em total}
stress-energy tensor only the metric coefficients $A$ and $B$ need
be known. In order to determine the metric nothing else than the total
stress-energy tensor is required. This property doesn't come
unexpected. It is exactly what is required from a universal theory
of gravitation that treats all forms of mass-energy equal.

However, whenever we care to distinguish between the "matter
contribution", and the "electromagnetic contribution", we must
know the electro-magnetic energy density $E^2/(8 \pi)$.

\section{Integration of the field equations for a charged holostar}

Whenever the sources of the gravitational field (matter and
electromagnetic) are known, the integration of the field equations
in order to obtain the metric is straight forward. In this paper
we are not interested in the general solution to the field
equations of a spherically symmetric charged system with an
arbitrary source-distribution, but rather in the charged extension
of the holostar solution.

The holostar-solution is characterized by the property $A B = 1$
throughout the whole space-time. I will assume that this property
remains valid for the charged case.\footnote{Note that $AB = 1$ is
fulfilled everywhere in the (charged) Reissner-Nordstr\"{o}m
space-time, except at the central singularity.} Setting $A B = 1$
allows the following simplification:

\begin{equation}
\frac{A'}{A} + \frac{B'}{B} = \left(\ln{A B}\right)' = 0
\end{equation}

from which

\begin{equation}
\rho + P_r = \rho_m + {P_r}_{m} = 0
\end{equation}

follows.

The field equations are reduced to the simple problem of solving
the following two equations:

\begin{equation}
\left( r B \right)' = 1-8 \pi r^2 \left( \rho_m + \frac{Q^2}{8 \pi
r^4}\right)
\end{equation}

\begin{equation}
8 \pi \left( {P_\theta}_m + \frac{Q^2}{8 \pi r^4}\right) =
\frac{B''}{2} + \frac{B}{r}
\end{equation}

They differ from the respective equations of the uncharged case
(with $A B = 1$) only in that the energy-density of the
electromagnetic field $E^2 / (8 \pi)$ is added to the mass-density
$\rho_m$ and to the tangential pressure ${P_\theta}_m$ of the
"matter" fields.

We are interested in a solution, where the sources of the fields
(electromagnetic and matter) are confined to a region $r \leq r_h$,
i.e. a solution that is situated in an exterior spherically
symmetric electro-vac space-time. The exterior solution therefore
must be given by the  well-known Reissner-Nordstr\"{o}m solution:

\begin{equation}
B_{e}(r) = \frac{1}{A_{e}(r)} = 1 - \frac{2 M}{r} +
\frac{Q^2}{r^2} = \left( 1-\frac{r_+}{r}\right)
\left(1-\frac{r_-}{r} \right)
\end{equation}

with

\begin{equation}
r_{\pm} = M \pm \sqrt{M^2 - Q^2}
\end{equation}

$Q$ is the total charge of the spherically symmetric
Reissner-Nordstr\"{o}m solution, $M$ its total gravitational mass.

The exterior electro-vac-solution has to be fitted to
the interior solution in such a way, that the metric remains
continuous at the boundary $r_h$ of the interior matter/charge
distribution.

We now have to construct an appropriate interior solution, for
which the uncharged holostar-solution is to serve as a guideline. The
(uncharged) holostar solution in \cite{petri/hol} is singled out
from all other possible interior solutions by the following
properties:

\begin{itemize}
\item $A(r < r_h) \propto r$

The radial interior metric coefficient $A$ is proportional to
$r$. Only such an interior metric leads to the prediction, that in
thermodynamic equilibrium the number of ultra-relativistic
particles is proportional to the proper surface area of the
object. I.e. the condition $A \propto r$ ensures, that the
interior solution is compatible with the holographic principle and
the Hawking entropy and -temperature laws.

\item $\rho = 1 / (8 \pi r^2)$

Related to the above condition is the requirement, that the
interior mass-energy density $\rho$ is fixed to $\rho = 1 / (8 \pi
r^2)$. If one assumes that the Einstein field equations with zero
cosmological constant are valid, this condition is equivalent to
the first.

\item $\int{T \sqrt{-g} dV} = M$

The improper integral over the trace of the stress-energy
tensor $T$, taken over the whole space-like volume, is exactly equal to
the gravitating mass $M$ of the holostar.

\item $\int{2 P_\theta \sqrt{-g} dV} \simeq M$

Alternatively or complementary to the above condition: The
"mass-energy content" of the membrane of the uncharged holostar, i.e. the
improper integral over the two non-zero principal pressure
components in the membrane, is (nearly) equal to its gravitating
mass $M$.

\item $P_\theta = 0$

The interior tangential pressure $P_\theta$ is exactly zero. In
combination with $\rho = - P_r$ this means, that the interior
matter has a string equation of state.

\end{itemize}

It is not a priori clear, how these properties are to be
generalized to the case of a charged (or a rotating) holostar.
Yet it is quite obvious that the essential requirement
responsible for the remarkable properties and self-consistency of
the holostar solution is, that the interior metric be subject to
the following condition:

\begin{equation} \label{eq:r/A:cond}
\left(\frac{r}{A}\right)' = 1 - 8 \pi r^2 \rho = 0
\end{equation}

i.e. $(rB)' = 0$ in the case of $A B = 1$.

Henceforth I assume that the condition $(r/A)' = 0$ holds for the
interior of the charged holostar. If this is so, the interior
metrics of the charged and uncharged holostar must be identical,
except possibly for a different integration constant $r_0$:

\begin{equation}
A = \frac{r}{r_0} = \frac{1}{B}
\end{equation}

With the above condition the total interior mass-energy density of
the charged holostar is uniquely determined:

\begin{equation}
\rho = \rho_m + \frac{E^2}{8 \pi} = \frac{1}{8 \pi r^2}
\end{equation}

The only difference to the uncharged case is, that the interior
mass density of the uncharged solution, $\rho$, must be replaced
by the total mass-energy density, which now consists of a
matter-contribution and an electromagnetic contribution.

It is reasonable to assume, that the matter-contribution and the
electromagnetic contribution to the interior mass-energy follow a
inverse square law in $r$, i.e. both are proportional to the total
mass-energy density with the same constant factor throughout the
whole interior. At least this is the simplest assumption
compatible with the requirement that the total mass-energy density
should scale with $1/r^2$. We arrive at the following ansatz for
the interior mass-density (matter-contribution):

\begin{equation}
\rho_m= \frac{c}{8 \pi r^2} (1-\theta)
\end{equation}

$\theta = \theta (r-r_h)$ is the Heavyside-step functional. To
save space the argument of the $\theta$-functional will not be
shown explicitly. Whenever a Heavyside $\theta$- or Dirac
$\delta$-functional is referenced in this paper, its argument will
always be given by ($r-r_h$). $c \leq 1$ is an arbitrary constant,
not to be confused with the velocity of light.

With the additional assumption $A B = 1$ we get $\rho_m + {P_r}_m
= 0$. This relation between mass-density and radial pressure is
not only true for the matter-part, but is also trivially fulfilled
for the electromagnetic contribution\footnote{A spherically
symmetric electric field always has $\rho = -P_r = P_\theta =
P_\varphi$.}, and therefore for the total mass-density and radial
pressure.

With the above ansatz for the matter-contribution the
energy-density of the electromagnetic field in the holostar's
interior must come out as:

\begin{equation} \label{eq:E2Ansatz}
\rho_{em} = \frac{E^2}{8 \pi}= \frac{1-c}{8 \pi r^2}
\end{equation}

$c$ will be less than 1, because the electromagnetic energy
contribution is always positive.

We now have to make an ansatz for the charge-distribution, i.e.
for the sources of the electromagnetic field, which is consistent
with equation (\ref{eq:E2Ansatz}). Principally there are two
physically distinct choices: The charge could be associated with
some of the holostar's interior matter, or with the membrane (or a
combination of both).

The first impulse would be to associate the charge entirely with
the membrane. Such an association appears to be indicated by the
membrane paradigm for a charged black hole \cite{Thorne/mem}.
Furthermore, if the holostar's charge could be attributed
exclusively to its boundary, one could view this as another independent
verification - or clarification - of the holographic principle.

If the total charge were assembled in the membrane there would be
no interior electromagnetic field. The interior energy density
will be "normal uncharged" matter. In this case the interior of a
charged holostar would be exactly equal to the interior of an
uncharged holostar.

Despite the apparent attractiveness of the above approach, in this
paper I take the position that the charge of the holostar must be
attributed to its interior particles and that the membrane is
essentially uncharged. This appears as the most natural choice:

We know that charged particles exist. We don't know of any
fundamental physical principle that forbids charged particles to
enter the holostar's interior. Therefore a charge-free interior
space-time, with all of the holostar's charge being placed at the
membrane, doesn't appear physically acceptable.

If the charge is due to the interior particles, cannot yet the
membrane carry a non-zero net-charge? In principle it could.
However, two arguments stand against such a possibility:

According to our present knowledge charge is always associated
with mass-energy. We don't know of any charged particle with zero
rest mass. If the membrane were charged, we should expect an
appreciable mass-energy density situated within the
membrane.\footnote{This will be even then the case, if charged
particles with zero rest-mass would exist. Such particles would
have to move with the speed of light within the membrane, and
therefore would carry energy. The lowest energy possible will be
comparable to the energy of a photon with a wavelength equal to
the proper circumference of the membrane. Therefore a membrane
containing charged particles would have to contain mass-energy.}
But the membrane of an uncharged holostar consists out of pure
tangential pressure and the mass-energy density $\rho$ within the
membrane is zero, at least in the uncharged case. There is
evidence, that this property of the membrane should also be valid
for the charged case:

If the membrane had a non-zero energy density, the metric wouldn't
be continuous. The metric coefficient $A$ is determined by the
integral over the total mass-energy density $\rho$, which consists
of a "matter term", $\rho_m$, and an electromagnetic term $E^2 /
(8 \pi)$. If the membrane carries a (finite) net charge $\Delta
Q$, the electromagnetic term to the total energy density $\rho$
will only contain a finite step-discontinuity.\footnote{If $Q$ is
the total charge of the interior particles and $\Delta Q \neq 0$
the non-zero net-charge of the membrane, the electric field at the
inner side of the membrane is given by $Q/r_h^2$, at its outside
by $(Q + \Delta Q)/r_h^2$. The energy density of the electric
field will jump at the membrane from $Q^2/(8 \pi r_h^4)$ to
$(Q+\Delta Q)^2 / (8 \pi r_h^4)$, which is finite.} The integral
over a finite step-discontinuity is continuous. There is no
problem with a discontinuity in the metric so far. The problem,
however, lies in the matter-term: We have assumed, that charge is
always associated with mass (or more generally with particles,
which have a non-zero energy even if their rest-mass is zero). If
this is true, any non-zero charge-density in the membrane,
$\rho_{c} \propto \Delta Q \delta(r-r_h)$, will require a non-zero
mass-energy density $\rho_m \propto \Delta Q \, (m/e) \,
\delta(r-r_h)$ in  the membrane, i.e. a non-zero
$\delta$-functional in the matter-term (m/e is the mass-to charge
ratio of the particles, which is very small in natural units for
all of the known particles). If the $\delta$-functional in the
matter term could be cancelled by a respective $\delta$-functional
in the electromagnetic term, we would be saved. But as was
established beforehand, the electromagnetic term at most contains
a $\theta$-functional. Therefore we have a non-cancellable
$\delta$-functional in the matter-term, which - when integrated -
produces a discontinuity in the metric coefficient $A$. This is
hardly acceptable. The continuity of the metric is essential for
the structure of general relativity.\footnote{The above argument
is based on the assumption, that the two-dimensional membrane is
"real", i.e. that it truly has no or negligible radial extension.
If the membrane is spread out over a radial coordinate region of
roughly Planck size, a non-zero net charge of the membrane might
be acceptable. In fact, even for a purely two-dimensional membrane
the relative change of the metric coefficients at the position of
the membrane is very small, if the charged membrane consists of
particles such as the electron or proton with small mass to charge
ratio $m/e$ in natural units. (For an electron $m/e \approx
10^{-22}$). If we find a discontinuity $\Delta B / B < 1$ in the
metric coefficients acceptable, we are led to the condition $Q <
(r_0 / 2) \, (e/m) \, \sqrt{r_h/r_0}$. In natural units $r_0/2
\approx 1$. Because of the large value of $e/m$ the above
condition is easily fulfilled for any "small" holostar with $r_h <
10^{40} r_0$. For large holostars we get a limit for the number of
charged elementary particles ($Q = e n_c$) in the membrane: $n_c <
1/\sqrt{2} \, (m_{Pl}/m) \, N^{1/4}$, with $N$ being the total
number of particles within the holostar. Therefore the number of
charged particles in the membrane grows only as the fourth root of
the total number of particles, or as the square-root of the
holostar's gravitational radius or mass.}

The other argument is based on the small ratio of gravitational to
electromagnetic force, at least for a large holostar. A
significantly charged membrane would produce a high electric
field, which will expel identically charged particles from the
membrane. This will almost instantly neutralize the membrane,
unless leaving the membrane requires a lot of work. This work
would have to be done against the gravitational field, whose
gradient is quite weak for a large holostar, especially in the
interior direction.\footnote{At the inside of the membrane of an
arbitrary holostar the proper gravitational acceleration $g_i$,
measured by an observer momentarily at rest with respect to the
membrane, is proportional to $1/r_h^{3/2}$, which is negligible
for a large holostar. The tidal forces are almost unnoticeable.
Although just outside of the membrane the proper gravitational
acceleration $g_o$ is large compared to the inside acceleration
($g_o \propto 1/r_h^{1/2}$), the value of $g_o$ can become
arbitrarily low for large holostars. The ratio of $g_o$ to $g_i$
at the membrane is given by $r_h/r_0-1$. Furthermore $g_o$
decreases very rapidly outside the membrane.} It is quite
improbable that the (gravitational) work required to leave the
membrane would turn out greater than the (electromagnetic) energy
that is released when the charged particles are expelled from the
membrane exclusively into the direction of the exterior
space-time.

A third argument might come from string theory. The membranes in
string theory (D-branes) usually have non-zero pressure, but zero
mass-energy density.\footnote{This has to do with the fact, that
the strings are merely attached to the membrane, but there
shouldn't be any string-segments lying within the membrane. In
string theory it is the string-segments which carry mass-energy,
whereas the string end-points only "carry" pressure.}

We therefore should associate the holostar's charge with its
interior particles. Charge is a relativistic invariant, therefore
we are forced to associate the charge density with the
number-density of the particles, and not with their
mass-density.\footnote{Contrary to charge, mass is not a
relativistic invariant. Therefore the ratio of charge- to
mass-density depends on the interior motion of the particles,
which is unsatisfactory.} The most natural assumption is, that
the ratio of charged particles to the total number of particles is
constant throughout the whole interior.

The problem with the above assumption is, that the number
densities of massive and massless particles are different in the
holostar. A choice has to made, whether to associate charge with
massive or with massless particles. But this choice has already
been made: So far as no massless charged particle is known to
exist, we should associate the charge with the number-density of
the massive particles, which scales as $n_m \propto 1/r^{5/2}$ if
no particles are created or destroyed.\footnote{This dependence is
derived in \cite{petri/hol}, where it is shown that the volume of
a geodesically moving spherical shell of massive particles evolves
with $r^{5/2}$. Assuming a constant number of particles in the
shell, the number-density evolves as the inverse volume. However,
the negative radial pressure in the holostar space-time leads to
particle creation/destruction, as the shell expands/contracts, so
that in general the assumption of a constant particle number in
the shell might not be valid. For charged particles, however, the
situation is different. Due to charge-conservation only uncharged
particles (or particle pairs) can be created by the pressure.
Charge conservation in combination with charge-quantization then
requires, that the difference of positively and negatively charged
particles must remain constant in the shell. The charge density
only depends on this difference, so that the net number-density of
charged particles always evolves as $n_c \propto n_+ - n_- \propto
1/r^{5/2}$.} We find:

\begin{equation} \label{eq:rho_em}
\rho_{c} = \frac{\kappa}{8 \pi r^2} \sqrt{\frac{r_0}{r}}(1-\theta)
\end{equation}

The factors of proportionality were arranged such, that $\kappa$
is a dimensionless constant. Its value will be determined later.

The energy density of the electromagnetic field can be derived
from the charge density. In the spherically symmetric case the
electromagnetic field strength $E(r)$ can be calculated just as in
the Newtonian case, via equations (\ref{eq:E(r)}, \ref{eq:Q(r)}).

Using equation (\ref{eq:rho_em}) for the charge density, we arrive
at the following result for the charge function $Q(r)$:

\begin{equation} \label{eq:Q}
Q(r) = \frac{\kappa}{2} r (1-\theta) + \frac{\kappa}{2} r_h \theta
\end{equation}

The total charge $Q(r)$ grows linearly with $r$ inside the
holostar (quite similar to the radial metric coefficient), and
remains constant outside.

From (\ref{eq:Q}) the electromagnetic field, and thus the field's
energy density follows:

\begin{equation} \label{eq:E2}
\frac{E^2(r)}{8 \pi} = \frac{\left(\frac{\kappa}{2} \right)^2}{8
\pi r^2} (1-\theta) + \frac{\left(\frac{\kappa}{2}\right)^2}{8 \pi
r^2} \frac{r_h^2}{r^2}\theta
\end{equation}

As required, the electromagnetic contribution to the interior
mass-energy density follows an inverse square law and is strictly
proportional to the matter contribution. This result is
non-trivial. If we had naively set the charge-density proportional
to the mass-density, we would have found a $1/r$-dependence of the
interior electromagnetic energy density, due to the factor
$\sqrt{A}$ in the proper volume element $dV$. Only by setting the
interior charge-density proportional to the {\em number-density}
of the massive particles, and by determining the charge $Q(r)$ by
integrating over the {\em proper} volume element, and not the
improper volume element $\widetilde{dV} = 4 \pi r^2 dr$, could
this result be achieved.

The total charge of the holostar, which can be measured by an
asymptotic observer in the exterior space-time by means of a
Gaussian flux integral, is given by:

\begin{equation} \label{eq:Qges}
Q = Q(r_h) = \frac{\kappa}{2} r_h
\end{equation}

This allows us to express $\kappa$ in terms of the total charge
$Q$ and the position of the membrane $r_h$:

\begin{equation} \label{eq:kappa}
\frac{\kappa}{2} = \frac{Q}{r_h}
\end{equation}

Therefore the electromagnetic and "matter" contributions to the
interior energy density are given in terms of the external
parameters $Q$ and $r_h$ as follows:

\begin{equation} \label{eq:rho:em:ext}
\rho_{em}(r \leq r_h) = \frac{Q^2}{r_h^2} \frac{1}{8 \pi r^2}
\end{equation}

and

\begin{equation} \label{eq:rho:m:ext}
\rho_{m}(r \leq r_h) = \left(1-\frac{Q^2}{r_h^2}\right) \frac{1}{8 \pi r^2}
\end{equation}

The respective values of the radial pressure are just the negative
of the respective energy densities, i.e. ${P_r}_m = -\rho_m$ and
${P_r}_{em} = -\rho_{em}$.

The metric must be continuous at the boundary, $r_h$. Therefore
the exterior Reissner Nordstr\"{o}m metric must match the interior
metric $B = 1/A = r_0/r$ at the position of the membrane.

This condition can be expressed as follows:

\begin{equation} \label{eq:Solve_rh}
\frac{r_0}{r_h} = 1 - \frac{2 M}{r_h} + \frac{Q^2}{r_h^2}
\end{equation}

which can be solved for $r_h$:

\begin{equation} \label{eq:rh}
{r_h}_{\pm} = M + \frac{r_0}{2} \pm \sqrt{(M+\frac{r_0}{2})^2
-Q^2}
\end{equation}

Equation (\ref{eq:rh}) only has a real valued solution for

\begin{equation} \label{eq:condM}
|Q| \leq M + \frac{r_0}{2}
\end{equation}

For all practical purposes ${r_h}_-$ in equation (\ref{eq:rh})
appears to be irrelevant:

Whenever $|Q| \leq M$ we are forced to take ${r_h}_+$: If we would
match the interior and exterior metric at ${r_h}_-$, the exterior
metric will undergo a sign change, i.e. $B$ will necessarily
become negative in some space-time region outside the
holostar.\footnote{It is easy to see, that for $|Q| \leq M$ the
(global) minimum of the exterior metric $B(r_{min}) = 1 - M^2/Q^2$
is negative. Furthermore it can be shown with a little bit of
algebra, that ${r_h}_- < r_{min}$, whenever $|Q| \leq M$.
Therefore if interior and exterior metrics are matched at
${r_h}_-$, the exterior metric will necessarily go through its
minimum, which is negative whenever $|Q| < M$.} This is not
desirable. We would like to construct a charged solution without
event horizon.

If interior and exterior metric are matched at ${r_h}_+$, the
membrane is the global minimum of the metric coefficient $B$,
whenever $|Q| \leq M$. As $B(r_h) = r_0/r_h$ is always positive,
$B$ will never undergo a sign change when the exterior and
interior metric are matched at ${r_h}_+$, and $|Q| \leq M$.

In contrast to the Reissner-Nordstr\"{o}m solution, however, $|Q|$
can be larger than $M$, albeit just by a small Planck sized
amount. Whenever the mass is in the range $M < |Q| \leq M + r_0/2$
it is possible to take either ${r_h}_+$ or ${r_h}_-$. There is no
problem with a sign-change in $B$, because interior and exterior
metric are always positive for $|Q| > M$ on the whole positive
$r$-axis.\footnote{The global minimum of the exterior metric is
positive for $|Q| > M$. The interior metric is always positive
(assuming $r_0 > 0$).}

Having two solutions is not desirable. In order to get a unique
solution in the range $M < |Q| \leq M+r_0/2$, the root in equation
(\ref{eq:rh}) should be set to $0$. Doing this we get $r_h = |Q| =
M+ r_0/2$, which corresponds the most extremely charged holostar
possible for a given mass $M$. The choice $|Q| = M+ r_0/2$ is
attractive for another reason: The interior and exterior metrics
match smoothly not only with respect to the metric-coefficients,
but also with respect to their first derivatives.

For an extremely charged holostar with $r_h = |Q| = M + r_0/2$ the
membrane is not positioned at the global minimum of $B$. The
minimum of $B$ lies outside the membrane, at $r_{min} = Q^2/M =
r_h(1+r_0/(2M))$. For a large holostar the position of the minimum
is almost identical with the position of the membrane. However,
whenever $M \ll r_0$, such as for an electron, $r_{min}$ can lie
several orders of magnitude outside the membrane. Note also, that
for an extremely charged holostar the membrane is situated at $r_h
= |Q|$, which is roughly one tenth of the planck length, if the
charge $Q$ is set equal to the electron charge $e$.\footnote{$e^2/
\hbar = \alpha \rightarrow e = \sqrt{\alpha} \sqrt{\hbar} =
\sqrt{\alpha} r_{Pl} \approx 0.085 r_{Pl}$}

If we know $r_0$, $M$ and $Q$, all quantities of the charged
holostar-solution, interior or exterior, are determined. The
gravitational mass $M$ and the charge $Q$ of the charged holostar
can be measured by an observer in the exterior space-time. The
only unknown quantity is $r_0$, which is expected to be very
small, roughly equal to the Planck-length.

\section{Some properties of the charged holostar solution}

In this section some properties of the charged holostar solution
are compiled for further reference.

The expressions for a charged holostar are very similar to the
respective expressions of a RN black hole, if we replace the
gravitational mass $M$ with $M + r_0/2$. Let us therefore define
the geometric mass

\begin{equation} \label{eq:Mg}
M_g = M+\frac{r_0}{2}
\end{equation}

and express all relations in terms of the geometric mass, whenever
appropriate.

\subsection{Metric}

The metric of the charged holostar solution can be expressed in
the following compact form:

\begin{equation} \label{eq:B:holo}
B = \frac{1}{A} = \frac{r_0}{r} (1-\theta) + \left(1-\frac{2M}{r}
+\frac{Q^2}{r^2}\right) \theta
\end{equation}

with

\begin{equation} \label{eq:rh:holo}
r_h = M_g + \sqrt{M_g^2 -Q^2}
\end{equation}

With the above expression for $r_h$, the position of the membrane
will always lie between the radius defined by the charge and the
radius defined by the mass of the holostar:

\begin{equation} \label{eq:rh:restraint}
|Q| \leq r_h \leq 2 M_g
\end{equation}

\subsection{Energy-density and radial pressure}

The total mass-density and radial pressure of the charged holostar
solution are given by:

\begin{equation} \label{eq:rho:tot}
\rho = \frac{1}{8 \pi r^2} (1-\theta) + \frac{Q^2}{8 \pi
r^4} \theta
\end{equation}

The total radial pressure is opposite to the mass-density:

\begin{equation} \label{eq:Pr:tot}
P_r = -\rho = -\frac{1}{8 \pi r^2} (1-\theta) -
\frac{Q^2}{8 \pi r^4} \theta
\end{equation}

Outside of the holostar the total energy-density and radial
pressure are solely due to the electromagnetic field. The interior
mass-density, which is equal to $1/(8 \pi r^2)$, splits into a
"electromagnetic" and a "matter" part. The same is true for the
interior radial pressure. The matter part carries a fraction $c$
of the total interior mass-density, $\rho_m = c \, \rho$, with $c$
given by:

\begin{equation} \label{eq:c}
c = 1-\frac{Q^2}{r_h^2} = 2 \left(1-\frac{M_g}{r_h}\right) =
\frac{2}{r_h} \sqrt{M_g^2 - Q^2}
\end{equation}

The other fraction, $Q^2/r_h^2$, is carried by the electromagnetic
field.

For $Q^2 = r_h^2 = M_g^2$, i.e. for an extremely charged holostar,
the total mass-density and the total radial pressure are
continuous at the position $r_h$ of the membrane.

\subsection{Tangential pressure}

The tangential pressure can be determined via equations
(\ref{eq:Pt}, \ref{eq:Pt:cont}). The easiest way is to determine
the total tangential pressure via equation $(\ref{eq:Pt:cont})$,
using the total energy-density and radial pressure given in
equations (\ref{eq:rho:tot}, \ref{eq:Pr:tot}). We find:

\begin{equation} \label{eq:Pt:holo}
P_\theta = \frac{c}{16 \pi r_h} \delta + \frac{Q^2}{8 \pi r^4} \theta
\end{equation}

with $c$ given by equation (\ref{eq:c}).

Similar to the uncharged holostar, the total interior tangential
pressure $P_\theta(r<r_h)$ is zero. There is a $\delta$-function
of tangential pressure at the membrane, which is non-zero, except
for an extremely charged holostar with $r_h = |Q| =
M+r_0/2$.\footnote{That the membrane vanishes for an extremely
charged holostar could already have been deduced from the fact,
that the total radial pressure (and energy-density) are continuous
at the boundary of the source distribution, $r_h$, for an
extremely charged holostar.} The tangential pressure outside of
the holostar is solely due to the exterior electromagnetic field.

Although the total interior tangential pressure is zero, the
respective electromagnetic and matter parts are generally
non-zero. They exactly cancel each other. The electromagnetic part
of the interior tangential pressure is positive and given by:

\begin{equation} \label{eq:Pt:i:em}
{P_\theta}_{em}(r < r_h) = \frac{1-c}{8 \pi r^2}
\end{equation}

The "matter contribution" to the interior tangential pressure is
always negative and given by:

\begin{equation} \label{eq:Pt:i:m}
{P_\theta}_m(r < r_h) = -\frac{1-c}{8 \pi r^2}
\end{equation}

For the special value $c=1/2$, i.e. when the interior
energy-density is distributed equally between the electromagnetic
field and the remaining "matter field", the matter contribution to
the interior stress-energy tensor is a genuine vacuum
stress-energy tensor, with $\rho = -P_r = -P_\theta = -
P_\varphi$.

For $c=1/2$ the position of the membrane is given by:

$$r_h = \frac{4}{3} M_g = \sqrt{2} |Q|$$

The value $c = 2/3$ is interesting as well. It is known that for
$Q^2 \geq 3/4 M^2$ the heat capacity of a Reissner-Nordstr\"{o}m
black hole becomes positive. For the charged holostar this happens
whenever $c \leq 2/3$, i.e. when the electromagnetic contribution
becomes larger than $1/3$ of the total (local) interior energy
density. In this case we have $r_h = 3/2 M = \sqrt{3} |Q|$.

\subsection{Integrated energy-densities}

For the uncharged holostar solution the "stress-energy" content of
the membrane was (almost) equal to the gravitating mass $M$ of the
holostar, whereas the integral over the trace of the stress-energy
tensor was exactly equal to the gravitating mass. We would like to
find out, whether the charged case differs from the uncharged case
in this respect.

The integrated energy-densities obtained in this section are
"normalized" to the position of the asymptotic exterior observer
at spatial infinity. An asymptotic observer will perform the usual
integral over the proper spatial volume element, which in
spherical coordinates is given by $dV = 4 \pi r^2 \sqrt{A} dr$.
However he will correct the (local) energy $dE = \rho dV$ of any
thin spherical shell by the gravitational redshift factor of the
shell with respect to his position, i.e. by $\sqrt{B}$. For the
charged as well as for the uncharged holostar we have $A B = 1$
throughout the whole space-time. Thus the red-shift corrected
proper integral over the energy-density is simply given by the
(improper) integral over the flat spherical volume element $4 \pi
r^2 dr$.\footnote{$E = \int{\sqrt{B} dE} = \int{\rho \sqrt{A B} 4
\pi r^2 dr} = \int{\rho 4 \pi r^2 dr}$, whenever $AB = 1$.}

The integral over the two tangential pressure components is given by:

$$\int_0^\infty{2 P_\theta 4 \pi r^2 dr}$$
$$ = \int_{r_h-\epsilon}^{r_h+\epsilon}{\frac{1-\frac{Q^2}{r_h^2}}{8
\pi r_h} \delta(r-r_h) 4 \pi r^2 dr} +
\int_{r_h}^\infty{\frac{Q^2}{4 \pi r^4} 4 \pi r^2 dr}$$

\begin{equation} \label{eq:Pt:int}
= \frac{r_h}{2}\left(1-\frac{Q^2}{r_h^2}\right)+\frac{Q^2}{2 r_h}
= \frac{r_h}{2}
\end{equation}

The integral consists of a contribution from the membrane and a
contribution from the exterior electromagnetic field. The
holostar's interior doesn't contribute, because the total
tangential pressure in the interior of the charged holostar is
zero.

The result of equation (\ref{eq:Pt:int}) is in some respect
similar to the uncharged case: For the uncharged holostar the
integral over the two tangential pressure components (which are
non-zero only in the membrane) also gives $r_h/2$. For an
uncharged holostar $r_h/2$ is nearly equal to its gravitational
mass, i.e. $r_h/2 = M(1+r_0/r_h)$. Contrary to the uncharged
holostar solution neither the integral over the membrane alone,
nor the integral over the whole space-time, yields a quantity that
is equal or proportional to the gravitating mass $M$ of the
charged holostar. In general for a charged holostar $M \neq
r_h/2$.

This is not overly disconcerting. Neither the energy-density nor
the principal pressures of a space-time are Lorentz-invariant quantities.
Their individual values depend on the coordinate system. But the
trace of the stress-energy tensor is coordinate-independent. We
therefore should be rather interested in how the integral over the trace
of the stress-energy tensor comes out for the charged holostar. We
find:

\begin{equation} \label{eq:T}
T = \rho - P_r - 2 P_\theta =
\frac{1}{4 \pi r^2}(1-\theta) - \frac{1-\frac{Q^2}{r_h^2}}{8 \pi
r_h} \delta
\end{equation}

Note that the trace over the stress-energy tensor gives a quite
accurate account on the location and "strength" of the sources of
the gravitational field. Outside of the holostar, where there are
no sources, $T$ is zero. Inside the holostar $T$ is proportional
to the energy-density of the fields associated with the interior
sources.

Integrating $T$ defined in equation (\ref{eq:T}) over the whole
space-time gives the following result:

\begin{equation} \label{eq:T:int}
\int_0^\infty{T 4 \pi r^2 dr} = r_h
-\frac{r_h}{2}\left(1-\frac{Q^2}{r_h^2}\right) = M+\frac{r_0}{2}
\end{equation}

The above integral doesn't include the negative point mass $M_0 =
-r_0/2$ at the center of the holostar. If we include the negative
point mass, or - which is the preferred procedure (see for example
\cite{petri/hol}) - if we start the integration not at the
unphysical region $r = 0$, but rather at $r = r_0/2$, the integral
over $T$, carried out over the whole physically meaningful
space-time, is exactly equal to the gravitating mass $M$ of the
charged holostar:

\begin{equation} \label{eq:T:int:phys}
\int{T \sqrt{A B} 4 \pi r^2 dr} = \int{T \sqrt{-g} dV} = M
\end{equation}

The total electromagnetic energy of the space-time, $E_{em}$, as
evaluated by an asymptotic observer at spatial infinity, is given
by the improper integral over the electromagnetic energy density
$E^2/(8 \pi)$. We find:

\begin{equation} \label{eq:Eem}
E_{em} = \int_0^\infty{\frac{E^2}{8 \pi} 4 \pi r^2 dr} =
\int_0^{r_h}{\frac{Q^2}{2 r_h^2} dr} +
\int_{r_h}^{\infty}{\frac{Q^2}{2 r^2} dr} = \frac{Q^2}{r_h}
\end{equation}

The electromagnetic energy splits into two terms, an interior part
and an exterior part, which are given by the above two integrals.
Both integrals are exactly equal. Therefore the electromagnetic
energy is distributed equally over the interior and the exterior
space-time. This is different from the "matter" part of the energy, which is
exclusively situated in the interior space-time.

\subsection{How are the external parameters $M$ and $Q$ related to the internal energy distribution?}

In this section we will analyze how the interior energy
distribution, i.e. the constant local ratio of electromagnetic
energy density, $E^2/(8 \pi)$, to the energy density of the
(remaining) matter, $\rho_m = c /(8 \pi r^2)$, relates to the
exterior parameters $M$ and $Q$.

For the following discussion it is convenient to define the
(modified) exterior mass to charge ratio:

\begin{equation} \label{eq:xi}
\xi = \frac{M_g}{Q} = \frac{M}{Q}(1+\frac{r_0}{2 M}) \in [1,
\infty)
\end{equation}

This ratio is always greater than 1. For an extreme holostar ($Q =
M_g$) the ratio is unity, whereas for an uncharged holostar $\xi
\rightarrow \infty$.

The following abbreviations are useful:

\begin{equation} \label{eq:alpha}
\kappa(\xi) = \left(1 \pm \sqrt{1-\frac{1}{\xi^2}} \right) \in
[1,2]
\end{equation}

\begin{equation} \label{eq:beta}
\lambda(\frac{r_0}{M}) = \left(1 + \frac{r_0}{2 M} \right) =
\frac{M_g}{M} \in (1,\infty)
\end{equation}

$\kappa$ is a quantity that more or less characterizes how heavily
charged the holostar is. For an extremely charged holostar we find
$\kappa = 1$, whereas an uncharged holostar is characterized by
$\kappa = 2$.

$\lambda$ is a measure how "heavy" the holostar is with respect to
the Planck-mass. Under the assumption that $r_0 \approx r_{Pl}$, a
large heavy holostar with $M \gg r_0$ has $\lambda \simeq 1$,
whereas a light holostar with $M \ll r_0$ has very high $\lambda$.

With $\lambda$ the modified mass-to-charge ratio $\xi$ can be
expressed in terms of the actual mass-to-charge ratio which can be
measured by the asymptotic observer:

\begin{equation} \label{eq:xi:b}
\xi = \frac{M}{Q} \lambda
\end{equation}

For a large holostar ($\lambda \simeq 1$) the modified
mass-to-charge ratio is nearly equal to the actual ratio.

With the above defined quantities the radial position of the
membrane, $r_h$, can be expressed as follows:

\begin{equation} \label{eq:rh_xi}
r_h = M \, \kappa \, \lambda = Q \, \xi \, \kappa
\end{equation}

For an extreme holostar ($\xi = \kappa = 1$) one gets $r_h = Q$:
The membrane is situated exactly at the radius defined by the
charge (in natural units). Whenever the extreme holostar is large
($\lambda \simeq 1$) we find $M \simeq Q$. This result is quite
similar to the relation $M = Q = r_+$ for the classical extreme
Reissner-Nordstr\"{o}m solution. ($r_+$ is the position of the
outer horizon of the Reissner-Nordstr\"{o}m solution).

For an uncharged or nearly uncharged holostar ($\kappa \rightarrow
2$) we find $r_h \approx 2 M$, analogous to the relation ($r_+ = 2
M$) for the Schwarzschild solution.

The coefficient $c$, which determines the relative contributions
of matter and electric field to the interior energy distribution,
can be expressed in terms of $\xi$ as well. We find:

\begin{equation} \label{eq:c_xhi}
c = 2 \frac{1-\xi}{(\xi \pm \sqrt{\xi^2 -1})} = 2
\frac{1-\sqrt{\kappa(2-\kappa)}}{\kappa}
\end{equation}

As could have been expected, for an extremely charged holostar
($\xi \rightarrow 1$ or $\kappa \rightarrow 2$) the
matter-contribution to the interior energy density becomes
arbitrary small with respect to the total energy density. An
extremely charged holostar therefore consists of pure
electromagnetic energy.

The total electromagnetic energy is given by:

\begin{equation} \label{eq:Eem2}
E_{em} = \frac{Q^2}{r_h} = \frac{|Q|}{\xi \kappa}
\end{equation}

For an extreme holostar ($\xi = \kappa = 1$) the total
electromagnetic energy is equal to the absolute value of the
charge $|Q|$.

With the above definitions the interior ratio of the "matter" part
of the energy density $\rho_m$ to the electromagnetic contribution
to the energy density $E^2 / (8 \pi)$ is given by:

\begin{equation} \label{eq:xi:em}
x_i = \frac{\rho_m}{\rho_{em}} = \frac{c}{1-c} = \frac{M^2}{Q^2} 2
\lambda^2 \kappa (\kappa-1)
\end{equation}

For a weakly charged holostar $\kappa \rightarrow 2$. In this case
the above formula is simplified:

\begin{equation} \label{eq:xi:weak}
x_i \simeq 4 \xi^2 =\left(\frac{2M}{Q}\right)^2 \lambda^2
=\left(\frac{2M}{Q}\right)^2 \left(1+\frac{r_0}{2 M} \right)^2
\end{equation}

If the holostar is large, i.e. $\lambda \simeq 1$, $x_i$ is
roughly given by:

\begin{equation} \label{eq:xi:weak:large}
x_i \simeq \left(\frac{2M}{Q}\right)^2
\end{equation}

The interior ratio of matter energy to electromagnetic energy is
proportional to the square of the exterior ratio of gravitating
mass to charge.

It is quite remarkable, that the quantities on the left hand side
of equation (\ref{eq:xi:weak:large}), which describe the local
distribution of energy in the holostar's interior, go in linearly,
whereas the quantities on the right hand side, which describe the
global distribution of mass and charge in the exterior space-time,
go in squared.\footnote{This might be interpreted as another hint,
that for the quantities measurable by an exterior observer, such
as $M$ and $Q$, not the quantities themselves, but rather their
squares are fundamental. In geometric units $c = G = 1$ both $M$
and $Q$ have dimensions of length (or mass/energy). Their squares
have dimension of area. Area has the same dimension as angular
momentum in geometric units, which is quantized in units of
$\hbar$. In quantum gravity area is quantized in terms of the
spin variables of the $SU(2)$ connection.

Note also, that the interior quantities, i.e. the
energy-densities, have dimensions of inverse area in natural units
$c=G=1$. Equation (\ref{eq:xi:weak:large}) relates the interior
energy-densities  (units: $1 / \hbar$) to the squares of the
exterior quantities (units: $\hbar$), which hints at some sort of
duality correspondence between these quantities.}

For an extremely charged holostar ($\kappa = 1$) the ratio $x_i$
goes to zero. This is as expected. The higher the charge, the more
the internal energy density will be dominated by the charge, i.e.
by electromagnetic energy, and not by the remaining matter. For an
extremely charged holostar the interior energy-density is entirely
electromagnetic.

So far the interior ratio of the energy-densities were expressed
in terms of the dimensionless ratio $Q/M$, or rather $Q/M_g$.
However, general relativity is a geometric theory. The mass $M$ is
not a genuine geometric, but rather a derived
quantity.\footnote{In the context of the holostar solution this
can be seen quite clearly in the appearance of $M_g = M + r_0/2$
instead of $M$ in various equations.} Yet for all black hole
solutions there is a one-to-one correspondence between the
gravitational mass $M$ and the horizon area $A$, which is related
to the entropy by the Hawking formula $S = A / (4 \hbar)$.
Therefore $M$ and $A$ are interchangeable. A similar result holds
for the holostar solution. From a geometric point of view it is
more natural to interpret $A$ as the fundamental variable. In this
respect it is quite remarkable, that the ratio of electro-magnetic
to total energy density can be expressed in a very simple way in
terms of the dimensionless ratio $Q^2/A$:

\begin{equation} \label{eq:ratio:em:tot}
\frac{\rho_{em}}{\rho} = \frac{Q^2}{r_h^2} = \frac{4 \pi Q^2}{A}
\end{equation}

with $A = 4 \pi r_h^2$. For an extremely charged holostar it is
easy to see that $Q = M_g$ so that $r_h = M_g = Q$ and therefore
$A = 4 \pi Q^2$. Thus for an extremely charged holostar the
interior electro-magnetic energy-density is identical to the total
energy-density, meaning that the interior energy-density is
consists exclusively out of electro-magnetic energy.

Clearly it is possible to relate $\rho_{em} / \rho$ to the
dimensionless ratio $Q/M$, or rather to the modified charge to
mass ratio $Q / M_g = 1 / \xi$. Using equation (\ref{eq:rh_xi})
for $r_h$ the above relation (\ref{eq:ratio:em:tot}) can be
transformed to:

\begin{equation}
\frac{\rho_{em}}{\rho} = \frac{1}{\xi^2 \kappa^2(\xi)} =
\frac{1}{\kappa^2(\xi)} \left( \frac{Q}{M_g}\right)^2
\end{equation}

The factor $1/\kappa^2(\xi)$ is only slightly dependent on the
modified charge to mass ratio. It increases monotonically with
$\xi = Q/M_g$ and varies between $1/4$ (for an uncharged holostar)
and $1$ (for an extremely charged holostar). For a moderately
charged holostar with $Q/M_g \ll 1$ we have:

\begin{equation}
\frac{\rho_{em}}{\rho} \rightarrow \frac{1}{4} \left(
\frac{Q}{M_g}\right)^2 \ll 1
\end{equation}

For an extremely charged holostar:

\begin{equation}
\frac{\rho_{em}}{\rho} \rightarrow \left( \frac{Q}{M_g}\right)^2
\rightarrow 1
\end{equation}

\subsection{\label{sec:extreme}Extremely charged holostars}

In this section I will briefly discuss the characteristic
properties of an extremely charged holostar. We find the following
relations:

\begin{equation}
r_h = Q = M + \frac{r_0}{2} = M_g
\end{equation}

\begin{equation}
r_0 = 2 (Q-M)
\end{equation}

For an extremely charged holostar with $M \ll Q$, a condition
which is very well fulfilled by all known elementary particles, we
find $r_0 \simeq 2Q \simeq 2 r_h$. In this particular case $r_0$
lies outside the membrane.

For an extreme holostar the membrane is not the global minimum of
the time coefficient of the metric $B$. The minimum lies at
$r_{min}$, which is given by:

\begin{equation} \label{eq:rmin}
r_{min} = r_h \lambda = Q \frac{M_g}{M}
\end{equation}

Whenever $\lambda = M_g/M$ is large, the minimum will lie very far
outside of the membrane.

The values of $B$ at the minimum and at the membrane are given by:

\begin{equation}
B(r_{min}) = 1 - \frac{1}{\lambda^2} = 1 - \frac{M^2}{M_g^2}
\end{equation}

\begin{equation}
B(r_h) = 2(1 - \frac{1}{\lambda}) = = 2(1 - \frac{M}{M_g})
\end{equation}

For a large holostar, with $\lambda \approx 1$ the time
coefficient of the metric $B$ is almost zero at the position of
the membrane. The same applies to the position of the minimum,
which is very close to the membrane. For a small holostar of
Planck mass or less, i.e. with $\lambda$ large, the situation is
quite different: $B(r_{min}) \simeq 1$ and $B(r_h) \simeq 2$.

The value $\lambda = 2$ is special. In this case $r_h = r_0 = Q$.
The gravitational mass is half the charge, i.e. $M = Q/2$. The
minimum of $B$ lies outside the membrane at $r_{min} = 2 Q$ with
$B(r_{min}) = 3/4$. The value of $B$ at the membrane is unity, i.e
$B(r_h) = 1$.

\section{Some remarks about the cosmic censorship hypothesis}

The condition $|Q| \leq M_g = M + r_0/2$ in equation
(\ref{eq:condM}) is very similar to the condition $M \geq |Q|$ for
the Reissner-Nordstr\"{o}m (RN) solution. The requirement, that
the charge of a self gravitating body shall never exceed its
gravitational mass, is postulated by the cosmic censorship
hypothesis: Whenever $Q > M$ the RN-solution exhibits a naked
singularity.

The only difference between the holostar-condition and the
condition derived from the cosmic-censorship hypothesis in a
RN-spacetime is, that the Planck-sized quantity $r_0/2$ is added
to the gravitational mass $M$ in the "holostar version" of this
condition.

The holostar condition $|Q| \leq M_g$ can be traced to a somewhat
different origin. The holostar has no naked
singularity.\footnote{In the purely classical treatment the
holostar-solution formally has a negative point-mass singularity
of roughly a Planck-mass at its center. However, this formal
"singularity" should be regarded as artifact of the classical
description, which is expected to break down at the Planck-scale,
anyway. The "singularity" is completely contained to a region of
Planck area/volume. Neither in the charged, nor in the uncharged
case does this formal negative mass "singularity" have any effect
outside of the Planck-region: The gravitational mass integrated
over a nearly Planck-volume defined by the fundamental length
scale $r_0$ (or $r_0/2$) is exactly zero, if the formal negative
point mass at the center is included in the integration. According
to the findings of loop quantum gravity (LQG), the geometry must
be considered to be discrete at the Planck-scale. It makes no
sense to probe a volume bounded by an area smaller than the
smallest area-eigenvalue of LQG. Therefore, taking the results of
LQG seriously, the holostar solution contains no singularity at
all. The effect of the negative point mass only becomes
noticeable, when we probe distances smaller than the Planck-scale,
or more accurately, when we probe into a space-time region bounded
by an area smaller than the smallest area-eigenvalue of LQG. This,
however, makes no sense.} There is no necessity to invoke the
cosmic censorship hypothesis.

Yet the holostar solution has an interior metric coefficient
$B$ , which falls off with $1/r$ and becomes nearly zero at the
surface of a large holostar: $B_{min} = r_0/r_h$. If the exterior
vacuum metric is to match the interior (non-vacuum) metric, the
exterior metric must approach zero at some coordinate value $r$ in
the exterior space-time. Therefore, for a large holostar the
exterior and interior metrics will be matched at a position very
close the event horizon of the exterior electro-vacuum metric. It is
obvious, that whenever the exterior vacuum metric has no event
horizon, it will be difficult, if not impossible, to match the
interior and exterior metrics.\footnote{In principle the time
coefficient of the exterior vacuum metric could come very close to zero
without ever reaching zero. If this is so, a match between
exterior vacuum and interior metric might be possible. However, for a
holostar of the size of the sun this would require that the
exterior vacuum metric $B_e$ should approach zero to the remarkable
accuracy of $B_e \approx 10^{-40}$, without actually hitting zero.
This is quite improbable.}

Whereas the cosmic censorship hypothesis postulates an event
horizon of the (global) vacuum space-time in order to conceal any
interior singularity, the holostar requires an (almost) event
horizon in the exterior vacuum space-time, in order to match the
interior and exterior metrics without unbelievable fine-tuning.

\section{\label{sec:r0}A rough lower bound for the "fundamental length" $r_0$}

So far I have merely assumed that $r_0$ is a quantity of roughly
Planck-size. This is a reasonable assumption. The Planck-length is
the only universal quantity with dimension of length that can be
constructed from first principles, and that at the same time is
independent from the internal workings of any particular
field-theory (except possibly gravity itself). In \cite{petri/hol}
$r_0$ has been determined to be roughly twice the Planck length
from cosmological data, i.e. $r_0 \approx 1.88 r_{Pl}.
$\footnote{$r_0$ was derived from the ratio of the average
matter-density of the universe to the fourth power of the
microwave-background temperature.} In \cite{petri/thermo} further
arguments were given, why $r_0$, or rather $r_0^2$, should be
regarded as a universal quantity, only moderately dependent on the
energy-scale. The arguments given in the above mentioned two
citations, however, refer to large holostars containing a
macroscopic number of interior particles. It is therefore
worthwhile to find out whether the assumption of a fundamental
length scale $r_0$ roughly equal to the Planck length is
compatible with what is known about the microscopic world.

Let us take the position, that $r_0$ is universal, i.e. the same -
or very nearly the same - quantity regardless of the size of a
self gravitating object.

Equation (\ref{eq:condM}) can be expressed as a condition
restraining the possible values for $r_0$:

\begin{equation}
r_0 \geq 2 (Q-M)
\end{equation}

Assuming that this relation must hold universally, if we apply
this relation to microscopic particles we find, that whenever the
mass of a particle is small with respect to its charge, the above
equation will serve as a lower bound on $r_0$. All of the known
charged fundamental particles have an extremely small mass
(expressed in natural units) with respect to their charge.
Therefore they are quite ideal candidates in order to determine a
lower bound for $r_0$. If we plug in the mass of the electron $m_e
\approx 10^{-23} \, r_{Pl}$ for $M$ and the electron charge $e =
\sqrt{\alpha} \, \sqrt{\hbar} \approx 0.1 \, r_{Pl}$ for $Q$, the
mass of the electron can be utterly neglected with respect to its
charge, because $m_e \ll e$ in natural units. We get

\begin{equation} \label{eq:r0:bound}
r_0 \geq 2 \sqrt{\alpha} \, r_{Pl} \approx 0.17 r_{Pl}
\end{equation}

in natural units (with $r_{Pl} = \sqrt{\hbar}$). $\alpha$ is the
fine-structure constant.\footnote{Note, that $\alpha$ depends on
the energy-scale, which hints that $r_0$, although postulated to
be universal in the sense that it should not depend on the size of
a self-gravitating object, might yet depend moderately on the
energy-scale via the running value of $\alpha$.}

If we take a muon, a tau or a W-boson (even a proton or any other
charged, but uncolored particle), the result is very much the
same. The masses of all of these particles are extremely small
with respect to their charge, which is always equal to (or a
multiple of) the charge of the electron. Note however, that all
known charged particles have non-zero spin. Therefore the
spherically symmetric charged holostar solution doesn't apply to
those particles. In the following section I will discuss some
properties of a possible extension of the charged holostar
solution to the spinning case, which is expected to give a better
estimate of $r_0$.

\section{\label{sec:electron}The electron as a charged, rotating holostar?}

The electron has a charge which is roughly one tenth of the
Planck-mass in natural units, whereas its gravitating mass is
roughly $4 \cdot 10^{-23}$ of the Planck mass.\footnote{Note that
the electron charge $e$ could have been used to define a system of
units similar to the Planck units. In "charge" units, the natural
length (or mass) scale is given by: $r_c = \sqrt{\alpha} \, r_{Pl}
\simeq 0.085 \, r_{Pl}$, which is roughly a factor of 11 smaller
than the Planck length. However, in contrast to Planck's constant
the effective charge depends on the energy scale.}

The spherically symmetric charged holostar allows solutions with
large charge, but negligible mass, in natural units. Whenever
$r_0/r_{Pl} < 2 \sqrt{\alpha} \simeq 0.17$ one can find a solution
with $Q = e$ and with arbitrarily small mass. In fact, the mass
could be identical zero. Such a solution, however, has zero
angular momentum.

No fundamental charged particle with spin 0 has been found so far.
This leads us to the assumption, that $r_0$ should be larger than
the bound in equation (\ref{eq:r0:bound}), i.e. larger than $0.17
\,  r_{Pl}$.\footnote{\label{fn:spin0}There are other arguments
which create some doubt whether a fundamental spin-0 particle
truly exists: According to the area-formula of quantum gravity a
spin-network state with a spin-0 link has zero area, and therefore
zero entropy. There is some evidence that the particles in the
holostar can be identified with the links of a large quantum
gravity spin-network (see section \ref{sec:spin:network}). A
particle with zero area and entropy, however, is difficult to
accept. Is this the death for the spin-0 s-electron, or even the
downfall of supersymmetry? Presumably not. Charge has a similar
effect as angular momentum in general relativity. Maybe loop
quantum gravity must be extended to incorporate charge into the
area formula, endowing a charged spin-0 particle with a non-zero
area.} In fact, if the running of the coupling constants is taken
into account, one gets $r_0 > 0.4 \, r_{Pl}$ for $\alpha \approx
1/25$ at the energy where the three coupling constants of the
Supersymmetric Standard Model are unified \cite{Wilczek/BeyondSM,
Roulet, Barger/supersymmetry}. If the holostar solution is to
describe charged elementary particles, angular momentum will have
to be included. As long as no solutions for a spinning holostar
are available, one must resort to approximate reasoning. It is
well known, that charged black holes have similar properties as
rotating black holes. By comparing the known formula for the
Kerr-Newman and Reissner-Nordstr\"{o}m solutions with the formula
for the charged holostar, we are led to extend equation
(\ref{eq:rh:holo}) in the following way:

\begin{equation} \label{eq:rh:j}
r_h = M_g + \sqrt{M_g^2 - Q^2 - \frac{J(J+1)}{M_g^2}}
\end{equation}

I.e the gravitating mass $M$ in the classical black hole formula
should be replaced by the "geometric mass" $M \rightarrow M_g =
M+r_0/2$ and the angular momentum $J^2 \rightarrow J(J+1)$.

In the classical limit, i.e. for $M \gg m_{Pl}$, the above formula
approximates the well known formula for the Kerr-Newman solution

\begin{equation}
r_+ = M+ \sqrt{M^2 - Q^2 - \frac{J^2}{M^2}}
\end{equation}

whereas in the limit $J=0$ we find the formula  (\ref{eq:rh:holo})
for the charged non-rotating holostar.

What are the conditions under which the holostar solution could be
an acceptable description for an elementary particle? Elementary
particles, such as the electron or neutrino, are completely
described by few parameters, such as their charge(s), their
angular momentum and mass. The same is true for black holes and
for the holostar solution. Furthermore all elementary particles of
the same kind are indistinguishable, very light (in natural
units), and the lightest particles of a certain class are
stable.\footnote{Or very long lived, with respect to the age of the
universe.} If we identify the holostar with a "particle" (be it
fundamental or composed), equation (\ref{eq:rh:j}) indicates that
several "particles" with the same charge and angular momentum, but
different masses are possible. The different mass-states can be
regarded as excitations. According to equation  (\ref{eq:rh:j})
the lightest "particle" with a given charge and angular momentum
is the extreme case for which the square-root in equation
(\ref{eq:rh:j}) vanishes.

If the (classical) electron is to be described by a holostar, it
is reasonable to assume that it will be an extreme holostar.
Besides that an extreme holostar has the smallest mass (for a
given charge and angular momentum), it has other properties which
make it a suitable candidate for an elementary particle: An
extreme holostar has no membrane. At least this is true for the
charged, non-rotating holostar. It should be true for the rotating
holostar as well. Second, not only the metric is continuous at the
boundary of an extreme holostar - if the membrane vanishes the
metric derivatives are continuous as well. Therefore an extreme
holostar should be particularly stable. From black hole physics we
have learned that extreme black holes have zero temperature. A
stable elementary particle quite certainly requires zero
temperature, otherwise its Hawking radiation would be devastating.
As the holostar is in many respects similar to a black hole, this
argument also points to the extreme holostar as the suitable
candidate for the description of an elementary particle.

For an extreme holostar the argument of the square-root in
equation (\ref{eq:rh:j}) is zero. This implies that $r_0$ can be
determined whenever the values $M$, $Q$ and $J$ of an extreme
rotating/charged holostar are known:

\begin{equation} \label{eq:r0:J}
\frac{r_0}{2} =
\sqrt{\frac{Q^2}{2}+\sqrt{\left(\frac{Q^2}{2}\right)^2+J(J+1)}} -
M
\end{equation}

We can plug in the values of the electron into equation
(\ref{eq:r0:J}):

\begin{equation} \label{eq:r0:electron}
\frac{r_0}{2 r_{Pl}} =
\sqrt{\frac{\alpha}{2}+\sqrt{\left(\frac{\alpha}{2}\right)^2+\frac{3}{4}}}
 - 4 \cdot 10^{-23} \simeq
\sqrt{\frac{\alpha}{2}+\sqrt{\left(\frac{\alpha}{2}\right)^2+\frac{3}{4}}}
\end{equation}

$\alpha$ is the fine structure constant, which
at $r \gg 10^{20} \, r_{Pl}$ is roughly $1/137$.

It is doubtable, whether the above formula will give an altogether
correct description of the electron. It would require that
classical general relativity were correct right down to the Planck
scale. Although we shouldn't expect highly accurate numerical
results, the above formula might allow us to make an order of
magnitude estimate, which - if we are lucky - could even be
correct on the tree or loop level.

Under the assumption that formula (\ref{eq:r0:electron}) is at
least approximately correct for an electron, the mass $M$ appears
as a very small correction with respect to the other quantities
$Q$, $J$ and $r_0$. This is as expected. An extreme "elementary"
holostar should have $M \approx 0$ in Planck units.

Neglecting the mass of the electron with respect to the other
quantities and using the actual value for the fine-structure
constant (at low energies and zero momentum transfer), we find:

\begin{equation} \label{eq:r0:electron:alpha}
r_0= 1.8651346 \, r_{Pl}
\end{equation}

Note that equation (\ref{eq:r0:electron}) can be interpreted as a
scaling law for the fundamental area $r_0^2$.\footnote{One might
rather consider $\pi r_0^2$ as the fundamental quantity. It is the
smallest possible surface area, that a holostar can have, as $r_h
= r_0/2$ for an extreme holostar and for given values of $J$, $Q$
and $M$ an extreme holostar has the smallest possible area of the
membrane.} With this interpretation the fundamental area depends
on the energy / distance scale via the coupling constant $\alpha$:

\begin{equation} \label{eq:r0^2:scalinglaw}
\frac{r_0^2(E)}{4 \hbar}=\frac{\alpha(E)}{2} +
\sqrt{\left(\frac{\alpha(E)}{2}\right)^2+\frac{3}{4}}
\end{equation}

Whenever $\alpha$ is small, i.e. in the low energy range, the term
under the root is dominated by the spin-term $3/4$. For
cosmological distances (low $E$) we can neglect $\alpha$ with
respect to $\sqrt{3/4}$. Setting $\alpha = 0$ we find:

\begin{equation} \label{eq:r0:electron:small}
\frac{r_0}{2 r_{Pl}} = \left( \frac{3}{4}\right)^{\frac{1}{4}}
\simeq 0.93
\end{equation}

Therefore $r_0 \simeq 1.86 \, r_{Pl}$.

Keep in mind that formula (\ref{eq:r0^2:scalinglaw}) only
incorporates the electro-magnetic coupling, i.e. can only be
expected to be valid at the low energy regime, where the only other
long range force, besides gravity, is the electro-magnetic
interaction. For higher energies the other coupling constants will
have to be included.

Quite interestingly $r = \sqrt[4]{3/4} \,  r_{Pl}$ is the minimum
eigenvalue of the length operator in loop quantum gravity as given
by \cite{Thiemann/length}. This points at a connection between the
(classical) holostar solution and loop quantum gravity.

In \cite{petri/hol} the scale parameter $r_0$ has been estimated
to be $r_0 \approx 1.88 \, r_{Pl}$ by comparing the average
mass-density in the universe, as derived from the WMAP data
\cite{WMAP/cosmologicalParameters}, to the microwave background
temperature. It is truly remarkable and quite likely not a
coincidence, that two quite different estimates of $r_0$, one in
the regime of elementary particles, the other in the regime of
cosmology, give practically the same result.\footnote{Note, that
both estimates refer to low energies, where $\alpha \approx 0$!}
We shouldn't become too enthusiastic, though: The errors in the
cosmological estimate are rather large, so that the almost exact
correspondence could be coincidental. Nevertheless, the above
finding can be interpreted as a very strong indication, that
$r_0^2$ is a truly universal quantity and that the interpretation
of equation (\ref{eq:r0^2:scalinglaw}) as scaling law is
essentially correct. This interpretation is further discussed and
enforced in \cite{petri/thermo}.

Note also, that from the viewpoint of loop quantum gravity (LQG)
it appears reasonable to assume that $r_0^2 = 4 \sqrt{3/4} \hbar$
nearly exactly at very large length scales. The smallest non-zero
area-value in quantum gravity is given by: $a_0 = 8 \pi \gamma
\hbar \sqrt{3/4}$. $\gamma$ is the so called Barbero-Immirzi
parameter, an undetermined parameter in LQG. If we compare the
smallest possible area-eigenvalue of loop quantum gravity to the
smallest possible area of a holostar, $A_0 = 4 \pi (r_0/2)^2$, we
can determine $\gamma$ by setting both areas equal. We find:
$\gamma = 1/2$.

The value of the fine-structure constant $\alpha$ depends on the
energy scale. If the predictions of the Grand unified theories
(GUTs) are correct, all of the coupling constants meet at a scale
of roughly $10^{3} \, r_{Pl}$, i.e. at an energy that is roughly a
factor of $10^{3}$ below the Planck energy. In most GUT-models
$\alpha_{GUT}$ is still small at this scale, usually around $1/25$ (see
for example \cite{Barger/supersymmetry, Wilczek/BeyondSM,
Roulet}).

Although the GUTs indicate that $\alpha_{GUT}$ remains small at
the Planck scale, it is instructive to find out how $r_0$ is
effected in the case of large $\alpha_{GUT}$. For $\alpha_{GUT} =
1$ equation (\ref{eq:r0:electron}) gives the following value for
$r_0$:

\begin{equation} \label{eq:r0:electron:1}
r_0 = \sqrt{6} r_{Pl} \simeq 2.45 r_{Pl}
\end{equation}

The estimate for large $\alpha_{GUT}$ is not much different from the
estimate of $r_0$ for small $\alpha$. We can be quite confident
that $r_0$ should lie in the range $1.86 < r_0 < 2.5$, i.e. be
roughly twice the Planck length, at any conceivable energy scale.

According to the discussion in \cite{petri/bh, petri/thermo} not
$r_0$, but rather its square (multiplied by $\pi$ or $4 \pi$)
should be considered as fundamental parameter. $r_0^2$ has the
dimension of area, which is equal to the dimension of action or
angular momentum in natural units $c=G=1$. In quantum physics
action and angular momentum are quantized in units of $\hbar/2$.
In loop quantum gravity the area operator is quantized (however
not in integer multiples of $\hbar/2$). Therefore it doesn't seem
unreasonable to assume that $r_0^2$ might be quantized as well and
possesses a non-zero minimum eigenvalue.

Any definite value for $r_0^2 = \beta \hbar$ requires a particular
value for $\alpha$, which is determined by solving the following
equation:

\begin{equation} \label{eq:r0:alpha2}
\frac{\beta}{4}=\frac{\alpha}{2}+\sqrt{\left(
\frac{\alpha}{2}\right)^2+\frac{3}{4}}
\end{equation}

For $r_0 = 2 r_{Pl}$, i.e. $\beta=4$ we get the following
"prediction" for $\alpha$ for a spin-half particle (at the energy-scale
defined by $r_0 = 2 r_{Pl}$):

\begin{equation} \label{eq:finestructure:const}
\alpha = \frac{1}{4}
\end{equation}

From equation (\ref{eq:rh:j}) we can see, that the membrane is
situated at $r_h = r_0/2$ for any "elementary" extreme holostar
with zero or negligible mass. Therefore the surface area of an
elementary, extreme holostar will be given by:

\begin{equation} \label{eq:A0}
\frac{A_0}{\hbar} = \frac{4 \pi r_h^2}{\hbar} = 4 \pi \left(
\frac{\alpha}{2} +
\sqrt{\left(\frac{\alpha}{2}\right)^2+\frac{3}{4}} \right) = \pi
\beta
\end{equation}

Its "cross-sectional area" $\sigma = \pi r_h^2$, which - according
to the Hawking formula is equal to its entropy $\sigma_S$ turns
out as:

\begin{equation} \label{eq:sigma}
\sigma_S = \frac{\sigma}{\hbar} = \frac{A_0}{4 \hbar} = \pi \left(
\frac{\alpha}{2} +
\sqrt{\left(\frac{\alpha}{2}\right)^2+\frac{3}{4}} \right) = \pi
\frac{\beta}{4}
\end{equation}

Note, that $\pi r_0^2 = \pi \beta \hbar$ is the smallest possible
surface area, that any holostar can have. For $\alpha = 1/4$ and
$j = 1/2$ we have $\beta = 4$, so that the area of such an
elementary holostar will be $A_0 = 4 \pi \hbar$ and its
"cross-sectional area" $\sigma_0 = A_0/4 = \pi \hbar$.

For $\alpha = 1/4$ and $j=0$ (which could be interpreted as a
s-electron\footnote{See however the discussion in footnote
\ref{fn:spin0}.}) we get the "prediction" $\beta=1$, i.e. $r_0 =
r_{Pl}$. In this case (no rotation) the area of the membrane would
be $\pi$ and the cross-sectional area $\pi/4$ in Planck units.

\section{\label{sec:spin:network}The holostar as a quantum gravity spin-network}

In this section I attempt to relate the area of a holostar with
the area-eigenvalues of quantum gravity. The main motivation
behind this undertaking is the observation, that the number of
punctures of a large loop quantum gravity (LQG) spin network state
is proportional to the area that is being "measured" by the
spin-network. Usually this area has been identified with the event
horizon of a large black hole. For the holostar we should identify
this area with the holostar's boundary area (i.e. with the
membrane).\footnote{This identification appears very much
preferable over the event horizon: Whereas the position of the
holostar's boundary is defined by matter, and thus respects
diffeomorphism invariance, this is definitely not the case for the
event horizon of a black hole, which is a fictitious "surface" in
vacuum, whose position cannot be determined by any local
measurement. In fact, in order to "know" the position of the event
horizon one has to have access to the whole space-time's future,
as the event horizon "moves" a-causally in anticipation of the
matter that will eventually pass it in the future.} But the
boundary area of the holostar is proportional to the number of its
interior particles with roughly the same factor of proportionality
as the number of punctures of the associated spin-network state.
Furthermore, both the particles in the holostar and the links of a
spin-network state carry spin\footnote{The reader might object,
that the two spin-variables describe different aspects of nature
and are not related. Although such a point of view is feasible,
there is not much predictive power in this viewpoint, whereas the
identification of the spins appears to yield consistent results.}
and for large spin-networks, as well as for large holostars, the
spin $1/2$ entities dominate. This suggests a connection between
the links of a large spin network state and the respective spins
of the relativistic interior particles of a holostar.

The area operator in loop quantum gravity has the following so
called "full" spectrum \cite{Ashtekar/area}:

\begin{equation}
A = 4 \pi \gamma \hbar \sum{\sqrt{2j_u(j_u+1) + 2 j_d(j_d+1) -
j_{t}(j_{t}+1)}}
\end{equation}

with

$$|j_u-j_d| \leq j_{t} \leq j_u + j_d$$

$\gamma$ is the Barbero-Immirzi parameter.

The $j_u$ and $j_d$ are positive half-integers and $j_{t}$ varies
in integer steps according to the triangle summation law for
angular momenta. $j_u$, $j_d$ and $j_t$ are the respective spin
quantum numbers for the links of quantum gravity spin network,
which "puncture" the surface that is just being measured and thus
endow the surface with a calculable quantum of area at each
puncture. The "full spectrum" of the area operator assumes, that
the nodes of a quantum gravity spin network can lie within the
surface. For any puncture the spins $j_u$ and $j_d$ label the
links that don't lie within the surface. $j_u$ is the spin quantum
number of the link on the "upper" side of the surface (assuming a
given orientation), $j_d$ labels the link on the "down" side, and
$j_t$ stands for the link that is tangential to the surface, i.e.
lies within the surface at the puncture. The spin-quantum numbers
of the links of a quantum gravity spin-network can only change at
the nodes. Only at a node three or more links can
join.\footnote{Whenever there are more than three links at a node,
the links can always be combined into at least three links, so the
three-valent nodes can be regarded as fundamental.} Links that lie
completely within the surface (such as closed loops) don't
contribute to its area. Therefore $j_t$ is only relevant, if there
is a node within the surface.

If one assumes that all of the nodes of the spin-network lie
outside the surface, we have $j_t = 0$ and $j_u$ = $j_d$. In this
case we get the "reduced" area spectrum, that was first derived by
Rovelli and Smolin \cite{Rovelli/Smolin}:

\begin{equation} \label{eq:areaspectrum:reduced}
A = 8 \pi \gamma \hbar \sum{\sqrt{j(j+1) }}
\end{equation}

The spin quantum number $j$ runs over all the punctures.

If a correspondence between the classical holostar solution and a
(large) quantum gravity spin-network is to be derived, an
extremely rotating holostar appears as the best suited starting
point:

The area of a large spin-network state is dominated by the spin
$1/2$ links \cite{Ashtekar/IsolatedHorizons2}. The number of
punctures of a large LQG-spin network therefore will be given by:

\begin{equation} \label{eq:N:spinnetwork}
N = \frac{A}{8 \pi \gamma \hbar} \sqrt{\frac{4}{3}}
\end{equation}

Let us denote by $J$ the "total spin" of the spin-network, i.e.
the sum of the spin quantum numbers of all of the links (over all
punctures). Due to the predominance of the links with spin $1/2$,
the "total spin" $J$ will be roughly equal to $N/2$ for large $N$.

\begin{equation}
J = \sum{j} \approx \frac{N}{2}
\end{equation}

If we replace $N$ by $J$ in equation (\ref{eq:N:spinnetwork}) we
get the following relation between area $A$ and the "total spin"
$J$:

\begin{equation}
A \approx 8 \pi J \hbar \gamma \sqrt{3}
\end{equation}

This looks very much like the relationship between area and
angular momentum of an extreme (maximally rotating) Kerr black
hole

\begin{equation}
A = 8 \pi J \hbar
\end{equation}

whenever $\gamma \approx 1/\sqrt{3}$ and if we identify the total
sum of the spins of the links with the angular momentum $J$ of the
black hole.

This relationship between the sum of the spins of large LQG spin
network state and the angular momentum area law for an extreme
Kerr black hole was already noted by Krasnov \cite{krasnov}.
Knowing that a spin-network can in principle consist of one single
link with macroscopic spin $J$ and area $A = 8 \pi \gamma J(J+1)$,
Krasnov argued, that the Immirzi parameter must be equal to or
greater than one, if spin network-states with macroscopic links
exist and their spin can be identified with the angular momentum
of a Kerr-black hole. If $\gamma < 1$ one could devise a
spin-network state with a few large links, whose "angular
momentum" exceeds that of an extreme Kerr black hole, i.e. would
correspond to the unphysical Kerr-black hole solution with a naked
singularity.

Here, however, we take the position that the links of the LQG
spin-network states are not combined into one large link, but
rather are dominated by the smallest links with $j = 1/2$ and each
link contributes individually to the area and entropy in the large
N limit. This seems a more natural choice, if the links are to be
identified with fundamental particles, which all have low spins
and definite entropies. In fact, all of the fundamental particles
of the Standard Model (not including the gauge bosons) are spin
1/2 fermions.

Lets make the argument more definite: Let us assume that any link
of a LQG spin network can be identified with an
(ultra-relativistic) interior particle of the holostar solution
and that the spin of the ultra-relativistic particle is equal to
the spin of the respective link. With this identification we
should be able to determine the Barbero-Immirzi parameter.

First, however, we have to establish whether the reduced area
formula of equation (\ref{eq:areaspectrum:reduced}) can be used.
There is evidence from the charged holostar solution, that an
extreme rotating holostar has no membrane. If there is no
membrane, there should be no nodes within the boundary surface of
an extreme holostar, which justifies using the reduced spectrum
(at least in the extreme case).

Second we have to know the exact number of punctures within the
boundary surface. Intuitively one expects, that the number of
punctures should be equal to the number of different spins, i.e.
the total number of the interior particles of the holostar.
However, we have to consider the possibility that there are links
in the spin-network (i.e. particles), that don't puncture the
boundary. The number of these links (particles) is expected to be
small: Every interior particle of an extremely rotating holostar
is believed to be aligned with respect to the rotation axis and
the sum of all commonly aligned spins of the interior particles is
(nearly) equal to the total exterior angular momentum of the
holostar (see the discussion in \cite{petri/thermo}). Therefore
each spin of an interior particle will be "visible" to the
exterior observer as a contribution to the exterior angular
momentum of the holostar. If the spin of an interior particle (to
be identified with a LQG-link) is to be "visible" for the observer
outside, its spin should "puncture" the boundary area. Therefore
the number of punctures should be (nearly) equal to the number of
interior particles of the holostar. Furthermore, the equations of
geodesic motion within the holostar space-time show, that any
geodesically moving particle must eventually cross the boundary
membrane, swinging back and forth between interior and exterior
space-time. So in a - at least semantically correct - sense, every
particle must "puncture" the membrane.

\subsection{\label{sec:Immirzi:fermionic}Determination of the Barbero-Immirzi parameter for a fermionic holostar}

We are now ready to determine the Barbero-Immirzi parameter. Let
us first consider the case of a holostar, whose (interior)
particle content is dominated by spin-1/2 fermions, which appears
to be the physically most relevant case.

The fermionic (spin 1/2) holostar is suggested by the current
results of loop quantum gravity, according to which the area of a
large spin-network state is dominated by spin $1/2$ links. The
identification of links with ultra-relativistic particles leads to
the assumption, that the interior relativistic particles of a
large holostar should consist predominantly out of spin-$1/2$
particles.\footnote{This is quite compatible with the result in
\cite{petri/thermo}, where it was shown that the mean spin quantum
number of the interior particles cannot be much larger than $1/2$,
otherwise the holostar would acquire a higher angular momentum
than an extreme rotating Kerr black hole, simply by aligning all
of the spins of its interior particles.} From the viewpoint of
holostar thermodynamics a holostar consisting only out of spin
$1/2$ fermions is the simplest possible model that
works.\footnote{There is no solution for a holostar in
thermodynamic equilibrium that consists exclusively out of bosons.
At least one fermionic species is required!}

With this assumption the area of the holostar according to the
loop quantum gravity area formula is given by:

\begin{equation}
\frac{A_{LQG}}{\hbar} = 8 \pi \gamma N \sqrt{3/4}
\end{equation}

where the number of punctures is set equal to the number of
ultra-relativistic spin $1/2$ particles within the holostar.

In \cite{petri/thermo} it has been shown, that the total number of
particles $N$ within the holostar is proportional to the area of
the membrane and is given by:

\begin{equation} \label{eq:N:holo}
\frac{A}{\hbar} =  4 \sigma N
\end{equation}

$\sigma$ is the (mean) entropy per particle. Its exact value
depends on the relative number of ultra-relativistic bosonic and
fermionic degrees of freedom and on the relation between the
chemical potentials.

If the area determined by loop quantum gravity is to be equal to
the area of the holostar given by equation (\ref{eq:N:holo}) the
Barbero-Immirzi parameter can be determined:

\begin{equation} \label{eq:Immirzi}
\gamma = \frac{\sigma} {2 \pi} \sqrt{\frac{4}{3}} =
\frac{\sigma}{\pi \sqrt{3}}
\end{equation}

Note that $\sigma$ is always slightly larger than $\pi$, at least
in the thermodynamic models discussed in \cite{petri/thermo}. If
the contribution of the bosonic degrees of freedom can be
neglected, i.e. $f_B = 0$, the entropy per particle at high
temperatures is given by $\sigma \simeq 3.3792$, so that the
numerical value of $\gamma$ turns out as:

\begin{equation} \label{eq:Immirzi:fermionic}
\gamma \simeq 0.621
\end{equation}

The value of the Barbero-Immirzi parameter in equation
(\ref{eq:Immirzi}) is larger than a factor of roughly $4.8$ than
the value that was determined in \cite{Ashtekar/IsolatedHorizons2}
by counting horizon-surface states:

\begin{equation} \label{eq:gamma:ashtekar}
\gamma_0 = \frac{\ln{2}} {\pi \sqrt{3}}
\end{equation}

The difference can be traced to the following reason. In
\cite{Ashtekar/IsolatedHorizons2} any spin $1/2$ link of a large
LQG spin-network is associated with an entropy of $\ln{2}$,
because there are two "area states" associated with a spin 1/2
variable.\footnote{The analysis in
\cite{Ashtekar/IsolatedHorizons2} is much more sophisticated. Yet
it turns out that the above statement approximates the true
picture fairly well in case of a large spin-network state. A more
accurate description is this: Any link with a given spin induces a
deficit angle on the surface punctured by the link. There are only
a small number of allowed deficit angles for a link with a given
spin. These deficit angles can be labelled by the "magnetic
quantum number" of the spin. A spin 1/2 link has two possible
deficit angles, $+ 2 \pi$ and $- 2 \pi$, corresponding to $m = \pm
1/2$, i.e. $\Delta \varphi = 4 \pi m$. So there are two different
"area-states" associated with a spin 1/2 link. A spin-1 link has
three choices of deficit angle, corresponding to $m = -1, 0, 1$
etc.} This is a perfectly reasonable assumption, which however is
based on the line of thought that the thermodynamics of a black
hole should be completely determined by the states of its event
horizon.\footnote{In my opinion there is one major draw-back in
the determination of the number of physically distinct horizon
surface states for a given macroscopic area according to
\cite{Ashtekar/IsolatedHorizons2}. The derivation in
\cite{Ashtekar/IsolatedHorizons2} assumes that all links (with the
same deficit angle) in any large spin-network state are
distinguishable. This is an assumption which is difficult to
accept. There is no "tag" on the links which would enable us to
discern any two links with the same spin and the same deficit
angle. In quantum theory we have learned, that all fundamental
objects with the same set of quantum numbers are indistinguishable
from each other. If there is no special marker (i.e. hidden
variable) on each spin 1/2 (or spin 1) link, how can any two
identical horizon surface patches punctured by a single spin 1/2
link and inducing the same deficit angle can be regarded as
distinguishable? Likewise there is no physical process
conceivable, which would allow us to keep track of every
fundamental surface patch of a macroscopic black hole in a way
such that all surface patches / punctures can be individualized.
The information required just to keep a record of the individual
"positions" of the surface patches constituting the event horizon
of a macroscopic black hole would require a second black hole with
at least the size of the black hole whose horizon surface states
we would like to keep track of. Not to forget that any
"measurement" of the "position" of a single surface patch accurate
enough to distinguish its position from that of its neighbors will
require a tremendous amount of energy: The neighboring surface
patches (punctures) are within a Planck distance. The energy
required to determine the "position" of just one surface patch
will necessarily disturb the whole configuration. In fact,
determining the "position" of just one single puncture with
Planck-accuracy will effectively "erase" the positions of roughly
$\sqrt{N}$ punctures, where $N$ is the total number of punctures
through the event horizon of the black hole.} In this work the
viewpoint is taken, that the entropy of a compact self gravitating
body doesn't correspond directly to some abstract, distinguishable
horizon surface states, but should be rather be determined from
the microscopic entropy of the principally indistinguishable
constituent particles within the interior space-time of the self
gravitating body. With this point of view the phase-space
available for the fundamental quanta of matter (=particles), which
are identified with fundamental quanta of geometry (=links of a
large spin-network state), is increased. Therefore any
spin-network link of a large spin-network should be associated
with an entropy $\sigma \approx \pi$, which not only derives its
value from the (two or more) horizon surface states, but from the
total interior phase space available to the "links". Equation
(\ref{eq:gamma:ashtekar}) then should be modified by replacing
$\ln{2}$ with $\sigma$, which is the mean thermodynamic entropy
per particle whose exact value can be determined from microscopic
statistical thermodynamics along the lines presented in
\cite{petri/thermo}.

It is remarkable, however, that the determination of the
Barbero-Immirzi parameter by counting the horizon surface states
of a quantum gravity spin-network and the semi-classical
determination of the Barbero-Immirzi parameter by counting
particle states in the interior phase space of the space-time
yields essentially the same result. Therefore we can be quite
confident, that classical general relativity, combined with
microscopic-statistical thermodynamics, truly arises in the large
N limit of quantum gravity.

With the Barbero-Immirzi parameter given by equation
(\ref{eq:Immirzi}) the area of the smallest spin-network state of
quantum gravity (one single half spin) can be determined. We find:

\begin{equation}
\frac{A_0}{\hbar} = 8 \pi \gamma \sqrt{j(j+1)} = 8
\frac{\sigma}{\sqrt{3}} \sqrt{\frac{3}{4}} = 4 \sigma
\end{equation}

We should be able to identify this area with the area and entropy
of the holostar's interior spin $1/2$ particles. The particle's
cross-sectional area $\sigma_0$, which is equal to its (intrinsic)
entropy according to the Hawking formula, turns out to be exactly
equal to the (mean thermodynamic) entropy per particle.

\begin{equation}
\frac{\sigma_0}{\hbar} = \frac{A_0}{4 \hbar} = \sigma
\end{equation}

Therefore the total entropy of a fermionic spin $1/2$ holostar is
nothing else than the sum over the Hawking-entropies of all of its
interior ultra-relativistic particles.\footnote{Note, however,
that the surface area of the particles, from which the entropy is
determined via the Hawking-formula, is not the area of the
(non-existent) event horizon of the particles, but rather the area
of their boundaries, i.e. the area of the membrane.}

This result is relevant with respect to the discussion in
\cite{petri/hol}. There it was shown, that in the high temperature
regime of the holostar the outward directed geodesic acceleration
and the inward directed pressure-induced acceleration exactly
cancel, whenever the cross-sectional area of the particles that
produce the pressure, expressed in Planck-units, is equal to the
mean entropy per particle $s$. It is quite remarkable, that loop
quantum gravity - with the numerical value of the Barbero-Immirzi
parameter given by equation (\ref{eq:Immirzi}) - appears to
deliver exactly the value required for the cross-sectional area of
the "pressure particles", in order that the holostar be truly
static in the high temperature regime.

There is reason to believe, that the fermionic holostar is a good
approximation to a realistic self gravitating object only at high
temperatures. However, if this is true, one should be able to
calculate the value of the "fine-structure constant" at the Planck
energy $\alpha_0$ via equation (\ref{eq:r0^2:scalinglaw}),
whenever the area of a fundamental spin 1/2 particle, $A_0$, is
known:

\begin{equation}
\frac{A_0}{\hbar} = 4 \pi \frac{r_0^2}{4 \hbar} = 4 \pi
\left(\frac{\alpha_0}{2} + \sqrt{\left(\frac{\alpha_0}{2}\right)^2
+ \frac{3}{4}}\right) = 4 \sigma
\end{equation}

This can be solved for $\alpha_0$:

\begin{equation} \label{eq:alpha:fermionic}
\alpha_0 = \frac{\sigma}{\pi} - \frac{3}{4} \frac{\pi}{\sigma}
\end{equation}

With $\sigma \simeq 3.3792$ we find:

\begin{equation}
\alpha_0 \simeq 0.378
\end{equation}

This is very close to the value $3/8$ predicted by $SU(5)$ for
$\sin^2{\theta_W}$ at the unification energy. The electromagnetic
coupling constant $\alpha$ is related to the $SU(2)$ coupling
constant $\alpha_2 = g^2/(4\pi)$ by the Weinberg-angle: $\alpha =
\alpha_2 \sin^2{\theta_W}$. With $\alpha_0 = 0.378$ and
$\sin^2{\theta_W} = 3/8$ at the unification energy, we get the
remarkable "prediction", that the unified $SU(2) / SU(3)$-coupling
constant $\alpha_{2/3}$ at the Planck energy should be nearly
unity, i.e.

\begin{equation}
\alpha_{GUT} \simeq \frac{0.378}{0.375} \simeq 1.009
\end{equation}

\subsection{\label{sec:Immirzi:supersymmetric}Immirzi parameter for a supersymmetric holostar}

Quite likely the assumptions in the previous section are too
simplistic. It might not be possible to neglect the bosonic
degrees of freedom completely in the holostar's
interior.\footnote{Note also, that in the simple thermodynamic
model discussed in \cite{petri/thermo} it has been assumed, that
all particles are ultra-relativistic. Whereas this is a reasonable
assumption at high temperatures, i.e. for small holostars, the
model in \cite{petri/thermo} will have to be extended to
incorporate more than two particle species, including massive
species to accurately describe the phenomena in the
low-temperature regime.}

There is another reason, why bosons might play an important role
in the holostar's interior: The holographic solution is based on
Einstein's equations with zero cosmological constant. A zero
cosmological constant, or rather a zero vacuum energy, can be
explained in theories with unbroken supersymmetry. This suggests,
that supersymmetry might play an important role within the
holostar, at least at high energies.

In this section I discuss a very simple "supersymmetric"
thermodynamic model of the holostar, which is characterized by an
equal number of fermionic and bosonic degrees of freedom of the
interior particles, i.e. $f_F = f_B$.

In \cite{petri/thermo} two possibilities for a supersymmetric
phase with $f_F = f_B$ were discussed:

The "normal" supersymmetric phase consists of a gas of equal
numbers of degrees of freedom for the ultra-relativistic fermions
and bosons, which are in thermal equilibrium with each other and
their anti-particles. It can be shown, that in the holostar
space-time the ultra-relativistic fermions must have a chemical
potential proportional to the radiation temperature $\mu_F = u_F
T$, where $u_F = \pi / \sqrt{3} \simeq 1.814$ is a positive
dimensionless constant (in units $k=1$). The anti-fermions have
the opposite chemical potential of the fermions, i.e.
$\overline{\mu_F} = - u_F T$ and all of the bosons have zero
chemical potential.

However, in \cite{petri/thermo} it was noted, that for equal
fermionic and bosonic degrees of freedom there exists a second,
"abnormal" supersymmetric phase, which has nearly identical
thermodynamic properties to the "normal" gas phase consisting
exclusively out of ultra-relativistic fermions and anti-fermions.
If one identifies the anti-fermions in the "normal" - exclusively
fermionic - gas phase with the bosons in the "abnormal"
supersymmetric phase, the thermodynamic properties of both phases
are nearly identical. The bosons in the "abnormal" supersymmetric
phase in a sense have "disguised" themselves as the anti-particles
of the fermions of the "normal" phase. This "abnormal"
supersymmetric phase is characterized by the property, that there
are only fermions and bosons (no anti-particles!) with equal
degrees of freedom, i.e. $f_F = f_B$, and that the chemical
potentials of fermions and bosons are related by $\mu_F + \mu_B =
0$. For this unconventional supersymmetric phase we have $u_F =
-u_B \simeq 1.353$.

Whenever one specifies the relation between the number of
fermionic and bosonic degrees of freedom, which in a
supersymmetric context is given by $f_F = f_B$, and the relations
between the chemical potentials ($\mu_F + \overline{\mu_F} = 0$
and $\mu_B = 0$ for the "normal" supersymmetric phase; $\mu_F +
\mu_B = 0$ and $\mu_F > 0$ for the "abnormal" supersymmetric
phase), the thermodynamic properties of the holostar solution are
completely determined. The mean entropy per particle, $\sigma$,
the relative number- and energy-densities of fermions and
anti-fermions to bosons and the respective entropies of the
fermions, $\sigma_F$, anti-fermions $\overline{\sigma_F}$ and
bosons, $\sigma_B$, can be read off from the tables given in
\cite{petri/thermo}.

The purpose of this section is to compare the properties derived
from the supersymmetric thermodynamic model of the holostar with
the results of loop quantum gravity. Most likely the right way to
do this is to choose the "normal" supersymmetric thermodynamic
model for such a comparison. The difficulty is, that one has to
know the mean spin of the fermions and bosons, which depends on
the fundamental supersymmetric particle group, or rather on the
relative numbers of fundamental particles with spins
$0,1/2,1,3/2,2$. So far there appears to be no theoretical
preference for a specific supersymmetric particle group. In such a
situation it might by safer to use a minimalistic approach, which
incorporates only the observed spin-varieties. The only
fundamental particles that are known to exist are spin 1/2
fermions (quarks and leptons) and spin-1 (gauge) bosons ($\gamma,
W_\pm, Z, g$). Quite interestingly, if one compares the "abnormal"
supersymmetric model with the LQG-spin-network description under
the assumption, that all the fermions have spin 1/2 and all the
bosons have spin-1, one gets an astoundingly consistent picture,
which seems worthwhile reporting, although it is yet not possible
to decide, whether this picture bares some deeper physical
significance.

For the "abnormal" supersymmetric model we can read off the mean
entropy per particle $\sigma$, the ratio of the number-densities
of fermions to bosons, $w$, and the entropies per fermion
$\sigma_F$ and boson $\sigma_B$ from the tables given in
\cite{petri/thermo}. We find:

\begin{equation} \label{eq:s}
\sigma = 3.37174
\end{equation}

\begin{equation} \label{eq:w}
w = \frac{N_F}{N_B} = 10.8031
\end{equation}

\begin{equation} \label{eq:sF}
\sigma_F = 3.1948
\end{equation}

\begin{equation} \label{eq:sB}
\sigma_B = 5.2835
\end{equation}

With the assumption that all of the fermions are spin-$1/2$
particles and all of the bosons carry spin-1 we can calculate the
surface area with the loop quantum gravity area formula:

\begin{equation} \label{eq:AQG:supersymmetric}
\frac{A_{QG}}{\hbar} = 8 \pi \gamma N \left(\frac{w}{1+w}
\sqrt{\frac{3}{4}} + \frac{1}{w+1}\sqrt{2}\right) =  0.913 \, (8
\pi \gamma N)
\end{equation}

The number of punctures $N$ is identified with the total number of
interior particles. This appears as the most reasonable
assumption. However, if the bosons preferentially assemble in the
membrane, as suggested in \cite{petri/thermo}, this assumption
might have to be modified.

By setting the areas of equation (\ref{eq:N:holo}) and equation
(\ref{eq:AQG:supersymmetric}) equal, we find the following value
for the Immirzi parameter:

\begin{equation}
\gamma = 0.5881
\end{equation}

We are now in the position to compare the entropies of the
fermions and bosons, as calculated by holostar thermodynamics and
as calculated via the loop quantum gravity area formula.

According to the LQG area formula, the entropy of a spin
$1/2$-fermion turns out as:

\begin{equation}
\sigma_F = 2 \pi \gamma \sqrt{3/4} = 3.200
\end{equation}

This is almost equal to the entropy per fermion determined from
holostar thermodynamics, $\sigma_F = 3.195$.

The loop quantum gravity result for the entropy per boson is given
by

\begin{equation}
\sigma_B = 2 \pi \gamma \sqrt{2} = 5.226
\end{equation}

which again is close to, but not equal to the value determined
from the thermodynamic model, $\sigma_B = 5.284$.

The very simple supersymmetric model discussed here appears to
give some quite consistent results. The entropies predicted by
holostar thermodynamics for fermions and bosons are within 1 \% of
the entropies we expect from the loop quantum gravity area formula
in combination with the Hawking entropy-area law. The
correspondence is not perfect. Further insights will be necessary
to resolve the discrepancies in a satisfying manner.

Note, that the basic assumption on which the "abnormal"
supersymmetric thermodynamic model is based, namely that at
ultra-high energies and temperatures the anti-fermionic degrees of
freedom become nearly indistinguishable - at least in a
thermodynamic sense - from the bosonic degrees of freedom, so far
has no serious theoretical justification. Therefore further
insights might very well lead us to the conclusion, that the
apparent correspondence between the "abnormal" supersymmetric
thermodynamic model and LQG was nothing more than a curious
numerical coincidence.

\section{Is entropy a (nearly) conserved quantity?}

As has been shown in \cite{petri/thermo} the entropy of a holostar
is proportional to its number of constituent particles. The
thermodynamic entropy per interior particle within a large
holostar, $\sigma$, is constant and slightly larger than $\pi$.
The value of $\sigma$ is quite independent of the specifics of the
thermodynamic model.

On the other hand, according to the discussion in the previous
sections it makes sense to identify the microscopic constituents
of the holostar (=particles) with "elementary" extreme holostars,
which are the smallest and lightest possible holostars for any
given value of particle angular momentum and charge. These
elementary extreme holostars have a non-zero boundary area $A_0 =
4 \pi r_h^2 = \pi r_0^2 = \pi \beta \hbar$. This boundary area
allows one to attribute an intrinsic entropy, $\sigma_i = A/(4
\hbar)$, to the particles via the Hawking entropy-area relation.

The boundary area can either be determined by the "semi-classical"
expression (\ref{eq:r0^2:scalinglaw}) for $r_0^2$ or by the LQG
area formula, if the value of the Barbero-Immirzi-parameter is
known.

Let us first discuss the intrinsic entropy of the particles
determined via the semi-classical expression for the fundamental
area $r_0^2$, given in equation (\ref{eq:r0^2:scalinglaw}). We
find:

$$\sigma_i = \pi \frac{\beta}{4} = \pi \left( \frac{\alpha}{2} + \sqrt{\left( \frac{\alpha}{2}\right)^2 + j(j+1)}\right)$$

All fundamental particles of the Standard Model are spin-1/2
fermions. For these the intrinsic entropy at low energies ($\alpha
\simeq 0$) is roughly $\sigma_i \simeq \pi \sqrt{3/4} \simeq 0.8
\,  \sigma \, - \, 0.84 \, \sigma$, with $\sigma$ being the
thermodynamic entropy, which lies in the range between $\sigma \in
[3.16, 3.38]$ for realistic thermodynamic models.

A particle well outside the gravitational radius of the holostar
is expected to have very low energy, so that its intrinsic entropy
will be quite a good lower bound for its real thermodynamic
entropy. Therefore we find, that the entropy of the whole system
is not very much affected, whether the particle is inside the
holostar, where it has a thermodynamic entropy of $\sigma$ in the
relatively narrow range given before, or whether it is outside the
holostar, where its entropy is at least $0.8 \, \sigma$.

Whenever a single elementary particle enters the holostar, the
holostar's entropy changes by the mean thermodynamic entropy per
particle $\sigma$, because the number of its constituent particles
has been increased by one.\footnote{The accretion process doesn't
necessarily have to be adiabatic. For non-adiabatic processes we
just have to wait long enough, until the holostar has attained
thermal equilibrium.} But the change in the (thermodynamic)
entropy of the holostar is nearly equal to the intrinsic entropy
of the particle that has entered the holostar, because its surface
area is roughly $A_0 \approx 4 \sigma \hbar$. Taking the intrinsic
(Hawking) entropy of the particle into account, the total entropy
before and after the merger is roughly equal. The same argument
applies to a process, where the holostar emits a particle (via
Hawking radiation) or when two holostars of arbitrary size collide
and join. Note, that the difference between the entropy before and
after the merger is largest, when a single particle enters a large
holostar. For large holostars the thermodynamic entropies of the
individual constituent particles are nearly equal, especially if
the holostars have nearly equal size. Therefore the larger any two
merging holostars become, the better their total entropies should
be conserved.

If we estimate the intrinsic entropy of a particle from the area
assigned to the particle by the loop-quantum gravity area formula,
we find that the entropy might not only be conserved
approximately, but appears to be conserved strictly for the
particular case of a fermionic holostar consisting only out of
spin-$1/2$ particles, when we choose $\gamma = \sigma / (\pi
\sqrt{3})$. However, one must keep in mind that $\gamma$ was
explicitly chosen such, that $\sigma_i = A_0/(4 \hbar) = \sigma$,
so strict entropy conservation is not a genuine prediction in this
particular case. Note also that $\gamma$ depends explicitly on
$\sigma$, which is a model-dependent quantity. One would rather
expect, that the Barbero-Immirzi-parameter is constant,
independent of the specifics of the thermodynamic model. Therefore
other values for the Barbero-Immirzi-parameter might be
contemplated, such as $\gamma = 0.5$, as suggested in section
\ref{sec:electron}. On the other hand, it has been speculated
whether $\gamma$ might undergo a - moderate - rescaling depending
on the energy scale. In fact, $\gamma$ is related to the
fundamental area, so if $r_0^2$ can be regarded as a running
quantity, the same might apply to $\gamma$. In such a case one
would expect $\gamma = 0.5$ at the Planck energy, becoming larger
for lower energies.

At the current state of knowledge the evidence is quite in favor
of approximate entropy conservation, but not strong enough to
postulate strict entropy conservation (which would be attractive
for theoretical reasons). Yet even approximate conservation of
entropy in general relativity is highly attractive, because it
forbids thought-experiments of the following kind: Imagine a large
black hole of galactic size or larger. Assemble a spherical shell
of matter with total mass equal to the mass of the imagined black
hole and place the shell close to its (imagined) gravitational
radius. The mass-density of the shell can be chosen arbitrary low.
The smaller the mass-density should be, the larger a black hole
has to be imagined. Without (approximate) entropy conservation
there appears to be no physical law that forbids us to arrange the
matter in the spherically symmetric shell in an orderly fashion,
so that its entropy can be chosen almost arbitrary small. Now let
the shell collapse under its own gravity. When it passes its own
gravitational radius, a black hole will form. When the
event-horizon (or the apparent horizon) forms, an enormous amount
of entropy will be generated almost on the spot. This is quite
unbelievable. No physical process should produce an arbitrary
large amount of entropy almost instantaneously.

If entropy is (approximately) conserved in general relativity, why
then do we live in an era where quite apparently $\Delta S > 0$,
i.e. the entropy increases over time? It has been speculated (see
for example \cite{Hawking/HistoryOfTime}) that this might have to
do with our present time-asymmetric situation, which is
characterized by the fact that we live in an expanding universe.
However, as far as I know no convincing explanation has been given
so far, why an expanding universe should be associated with
$\Delta S > 0$, whereas in a phase of contraction the entropy
should shrink. In the holostar there appears to be some sort of
rudimentary explanation: In the low density regions of a holostar,
the number of photons with respect to the number of massive
particles increases over time, if the motion of the outmoving
massive particles is nearly geodesical, in fact
$\overline{n}_\gamma / \overline{n}_m \propto \sqrt{\tau}$ (see
\cite{petri/hol}). If we define the co-moving volume of the
geodesically moving massive particles as reference, the entropy
within this volume increases throughout the expansion, because
there will be more and more photons per nucleon. Furthermore,
according to the discussion in \cite{petri/hol} the negative
pressure in the holostar space-time has the effect to create new
(massive) particles within any expanding volume, which further
increases the entropy within this volume. For particles moving
inward on a geodesic trajectory (i.e. the time-reversed situation)
the number of particles in the co-moving volume shrinks, so
contraction is associated with $\Delta S < 0$. In \cite{petri/hol}
a more detailed discussion of global and local entropy- and
energy-conservation is given. There it will be shown, that the
total entropy and energy within the holostar space-time is
conserved. However, the total energy and entropy increase within
the local Hubble-volume of an outward moving observer, whereas
both quantities decrease for an inward moving observer. Increase
and decrease are exactly opposite to each other at every interior
space-time point, so in toto entropy and energy are conserved.

\section{Discussion}

The holostar solution has been generalized to the spherically
symmetric charged case. Many of the characteristic properties of
the uncharged holostar also apply for the charged case.

The charged holostar solution possesses a remarkable degree of
self-consistency. The interior charge distribution, from which the
total charge can be determined by proper integration, is
proportional to the number density of its interior massive
particles. This is not self-evident and depends on the subtle
interplay between the metric and the interior energy density, from
which the metric is derived. The exterior metric and electric
field are identical to the Reissner-Nordstr\"{o}m electro-vac
solution. Similar to the uncharged holostar solution, the total
gravitational mass $M$ of the space-time can be derived from a
proper integral over the trace of the stress-energy-tensor, a
Lorentz invariant quantity.

The charged holostar solution allows us to make some predictions,
which are quite in accord with the expectations from black hole
physics: A charged holostar solution with $|Q| \gg M$ is not
possible. Whereas in the case of electro-vac space-times solutions
with $|Q| > M$ are ruled out by the cosmic censorship hypothesis,
a holostar solution with $Q$ much higher than $M$ in natural units
simply doesn't exist.

Extremely charged holostars are of particular interest. They
consist exclusively out of electro-magnetic energy. The energy
density is continuous and the metric is differentiable continuous
across the boundary. Charged extreme holostar solutions with
negligible mass (in natural units) are among the possible
solutions. Those solutions have many features in common with
elementary particles.

By extending the formula describing an extremely charged holostar
to the case of angular momentum (and charge), the scale parameter
$r_0$ has been estimated. $r_0^2$ amounts to roughly four times
the Planck area, which is quite in agreement with the experimental
determination of $r_0^2$ in a macroscopic (cosmological) context.
There is a theoretical preference for a value of $r_0^2 \simeq 4
\sqrt{3/4}$ at macroscopic scales and $r_0^2 \simeq 4 \sigma /
\pi$ at the Planck scale, where $\sigma > \simeq \pi$ is the mean
entropy per ultra-relativistic particle determined by holostar
thermodynamics (see \cite{petri/thermo}). $r_0/2$ is the radial
coordinate value of the membrane of an elementary extreme holostar
of (nearly) zero mass.

If the holostar solution is identified with a large quantum
gravity spin-network state, the Barbero-Immirzi parameter $\gamma$
can be determined. We find $\gamma = \sigma / (\pi \sqrt{3})$,
when the ultra-relativistic particles of the holostar solution are
identified with the links of a quantum gravity spin-network state.
$\gamma$ is roughly a factor of $4.8$ higher than the value
determined in \cite{Ashtekar/IsolatedHorizons2} by counting
horizon surface states. The exact value of $\gamma$ depends on the
particle content of the thermodynamic model (i.e. the number of
fermionic and bosonic degrees of freedom with different spins) and
on the respective relations between the chemical potentials.

The identification of an elementary extreme holostar of minimal
mass with an elementary particle leads to the conjecture, that the
entropy of a self-gravitating system in general relativity should
be (approximately) conserved. The observation, that entropy
increases with cosmological time appears as a consequence of our
current time-asymmetric situation, i.e. that we live in the
expanding sector of the universe. An argument is given, why
expansion should lead to $\Delta S > 0$ and contraction to $\Delta
S < 0$.

A major topic of future research will be the search for a rotating
holostar, which should enable us to resolve many of the yet open
questions addressed in this paper and to put the preliminary
results presented here on a firmer basis. Hopefully the charged
holostar solution will provide some theoretical guidance to find a
charged/rotating solution.


\end{document}